\documentclass{article}
\usepackage[utf8]{inputenc}

\RequirePackage[T1]{fontenc} \RequirePackage[tt=false, type1=true]{libertine} \RequirePackage[varqu]{zi4} \RequirePackage[libertine]{newtxmath}

\usepackage{geometry}
\usepackage{hyperref}
\usepackage{mathrsfs}
\usepackage{tikz}
\usetikzlibrary{arrows.meta}
\usepackage{overpic}
\usepackage{pict2e}
\usepackage{dsfont}
\usepackage[most]{tcolorbox}
\tcbuselibrary{listings}
\tcbuselibrary{breakable}
\usepackage{listings}
\usepackage[outline]{contour}
\usepackage{standalone}
\usepackage[normalem]{ulem}
\usepackage{booktabs}  
\usepackage{array}     
\usepackage[table,xcdraw]{xcolor} 
\usepackage{color, colortbl}
\usepackage{microtype}      
\usepackage[capitalize]{cleveref}
\usepackage{xspace}         
\usepackage{mathtools}      
\usepackage{nicefrac,xfrac} 
\usepackage{mathtools}

\usepackage{stmaryrd}  
\usepackage{newfloat}  
\usepackage{savesym}
\usepackage{algorithm}
\usepackage{algorithmicx}
\usepackage[noend]{algpseudocode} 
\algrenewcommand\textproc{}
\usepackage{graphicx}
\usepackage{grffile}  
\usepackage{soul}
\usepackage{wrapfig}
\usepackage{bbm}
\usepackage{flushend}
\usepackage{tabularx}
\usepackage{ctable}            
\usepackage[a]{esvect}           
\usepackage{multirow}

\usepackage{wrapfig}
\usepackage{textcomp}    
\usepackage{enumitem}    
\usepackage{natbib}

\hypersetup{
    colorlinks=true,
    linkcolor=blue,
    filecolor=magenta,      
    urlcolor=blue,
    pdftitle={Noise lab},
    pdfpagemode=FullScreen,
    }

\newcommand{\eg}{e.g.,\xspace}
\newcommand{\ie}{i.e.,\xspace}

\newcommand{\expectation}{\mathbb{E}}

\newcommand{\domain}{\Omega}
\newcommand{\sample}{\mathbf{x}}
\newcommand{\integral}{I}
\newcommand{\samplecount}{N}
\newcommand{\mce}{\integral_\samplecount}

\newcommand{\timestep}{\tau}
\newcommand{\noiseterm}{\xi}
\newcommand{\normaldistribution}{\mathcal{N}}
\newcommand{\temperature}{T}

\newcommand{\wienerprocess}{W}
\newcommand{\driftterm}{\mu}
\newcommand{\diffusionterm}{\sigma}

\newcommand{\generalcoord}{q}
\newcommand{\hamiltonian}{H}
\newcommand{\generalmomentum}{m}
\newcommand{\potentialenergy}{U}
\newcommand{\kineticenergy}{K}
\newcommand{\energy}{f}
\newcommand{\normalization}{Z}

\newcommand{\param}{\theta}
\newcommand{\loss}{\mathcal{L}}
\newcommand{\learningrate}{\eta}

\newcommand{\encoderparams}{\phi}
\newcommand{\decoderparams}{\theta}

\newcommand{\score}{\mathbf{s}_\theta}
\newcommand{\targetdistribution}{p}
\newcommand{\decodertrue}{p}
\newcommand{\datadistribution}{p}
\newcommand{\encodertrue}{q}
\newcommand{\encoder}{q_\encoderparams}
\newcommand{\decoder}{p_\decoderparams}
\newcommand{\model}{p_\theta}
\newcommand{\latentvar}{\mathbf{z}}

\newcommand{\kldivergence}{D_{KL}}
\newcommand{\totaltime}{T}
\newcommand{\gradientx}{\nabla_\sample}

\newcommand{\pathspace}{\domain}
\newcommand{\set}[1]{\left\{#1\right\}}
\newcommand{\vx}{\mathbf{x}}
\newcommand{\vy}{\mathbf{y}}
\newcommand{\vu}{\mathbf{u}}
\newcommand{\where}{\;\middle\vert\;}
\newcommand{\surfaces}{\mathcal M}

\newcommand{\IGNORE}[1]{}   
\newcommand{\wj}[1]{\IGNORE{{\color{blue}#1}}}
\newcommand{\gurprit}[1]{\IGNORE{\textcolor[rgb]{1,0.27,0}{\textit{[Gurprit: #1]}}}}

\DeclareGraphicsExtensions{.png,.jpg,.pdf,.ai,.psd}
\begin{document}

\begin{titlepage}
    \begin{center}
    {\Huge{\bf MCMC: Bridging Rendering, Optimization, and Generative AI}}\\*[3mm]
    \vspace{0.5cm}
    \vspace{0.5cm}

    {\large Gurprit Singh$^1$, Wenzel Jakob$^2$}\\
    \vspace{0.5cm}
    {\large $^1$Max Planck Institute for Informatics, Germany, $^2$EPFL, Switzerland}\\
    \vspace{0.5cm}
    {\large SIGGRAPH Asia 2024 Course notes}\\
    \vspace{0.5cm}
    
    \end{center}
    
    \paragraph{Abstract.} 
    Generative artificial intelligence (AI) has made unprecedented advances in vision language models over the past two years. 
    These advances are largely due to diffusion-based generative models, which are very stable and simple to train.    
    These diffusion models are tasked to learn the underlying unknown distribution of the training data samples. 
    During the generative process, new samples (images) are generated from this unknown high-dimensional distribution. 
    Markov Chain Monte Carlo (MCMC) methods are particularly effective in drawing samples from complex, high-dimensional distributions. This makes
    MCMC methods an integral component for both the training and sampling phases of these models, ensuring accurate sample generation.
    
    Gradient-based optimization is at the core of modern generative models. The update step during the optimization forms a Markov chain where the new update depends only on the current state. 
    This allows exploration of the parameter space in a memoryless manner, thus combining the benefits of gradient-based optimization and MCMC sampling.
    MCMC methods have shown an equally important role in physically based rendering where complex light paths are otherwise quite challenging to sample from simple importance sampling techniques. 
    
    A lot of research is dedicated towards bringing physical realism to samples (images) generated from diffusion-based generative models in a data-driven manner, however, a unified framework connecting these techniques is still missing.   
    In this course, we take the first steps toward understanding each of these components and exploring how MCMC could potentially serve as a bridge, linking these closely related areas of research. 
    Our course aims to provide necessary theoretical and practical tools to guide students, researchers and practitioners towards the common goal of \emph{generative physically based rendering}.
    All Jupyter notebooks with demonstrations associated to this tutorial can be found on our project webpage~\href{https://sinbag.github.io/mcmc/}{https://sinbag.github.io/mcmc/}.

\end{titlepage}

\tableofcontents

\newpage

\section{Overview}

\begin{figure*}[!h]
    \centering
    \includegraphics[width=\linewidth]{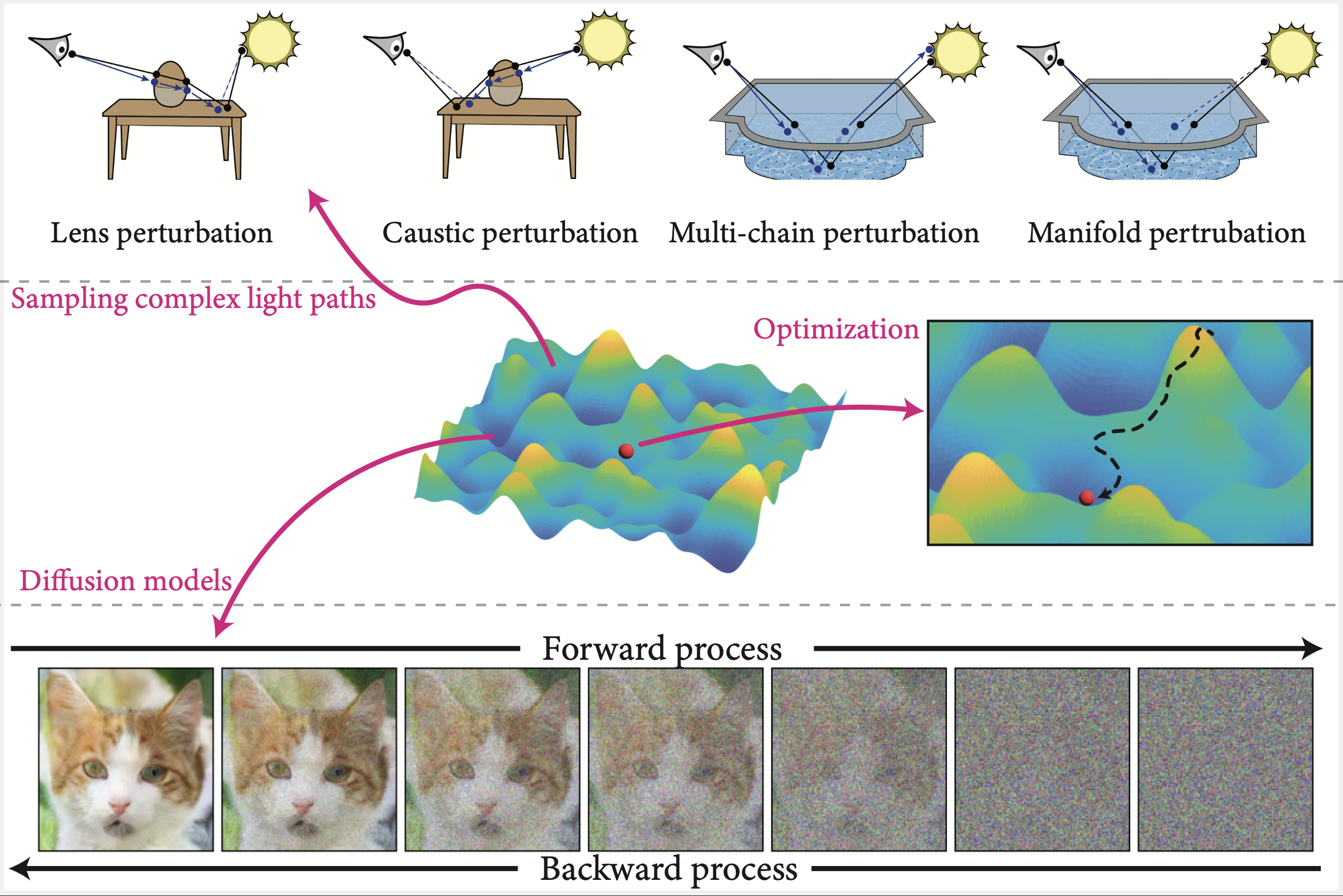}
    \caption{
    This course provides an overview of MCMC methods, which are powerful tools for sampling from complex distributions. An example of such a distribution is shown in the middle row (Part II). These complex distributions are common in physically based rendering and generative modeling. Additionally, we will review the impact of MCMC methods on gradient-based optimization techniques, which typically aim to find good local (or global) minima in the optimization landscape (inset on the middle-right).
    }
    \label{fig:overview}
\end{figure*}

MCMC methods are powerful tools for sampling from complex, high-dimensional probability distributions, which are ubiquitous in modern computational problems. Whether you are working with Bayesian inference in statistics, machine learning, or any field involving probabilistic models, MCMC provides a robust framework for gaining insights and making accurate predictions.

In this course, we introduce the terminology associated to probabilistic models. We start from the theoretical foundations essential to establish the ground work for Markov chains~\cref{sec:theoretical_background}. We introduce stochastic differential equations (SDEs) that are essential to describe stochastic systems evolving over \emph{time}. Brownian motion is one such example. We later on discuss Langevein and Hamiltonian dynamics which are capable of exploring the anisotropic regions far more efficiently. Markov chains are a memoryless way to sample paths described by the SDEs. In~\cref{sec:mcmc_rendering}, we discuss various Markov chain Monte Carlo (MCMC) sampling methods with direct applications to physically based rendering. In~\cref{sec:mcmc_optimization}, we introduce stochastic gradient descent (SGD) optimization algorithm which is at the core of all machine learning tasks. 
We show that SGD update step can be seen as a Markov chain and more advanced MCMC sampling methods can be employed to better explore the optimization manifolds.
Lastly, in~\cref{mcmc_generative} we study variational autoencoders, how they are driven by evidence lower bound and their connection to variational diffusion models. We then discuss energy-based models which, although slow to train, provides a lot of flexibility.  We conclude with the exciting research directions that can follow from this course.

\section{Theoretical background}
\label{sec:theoretical_background}

Markov Chains are mathematical models used to describe systems that transition from one state to another, where the 
probability of each transition depends only on the current state (this property is known as the "memoryless" property or Markov property). The system moves between discrete states with given transition probabilities. Markov Chains are often represented as 
sequences of random variables.
%
\vspace{-20pt}
\begin{figure*}[h]
    \centering
    \includegraphics[width=\textwidth]{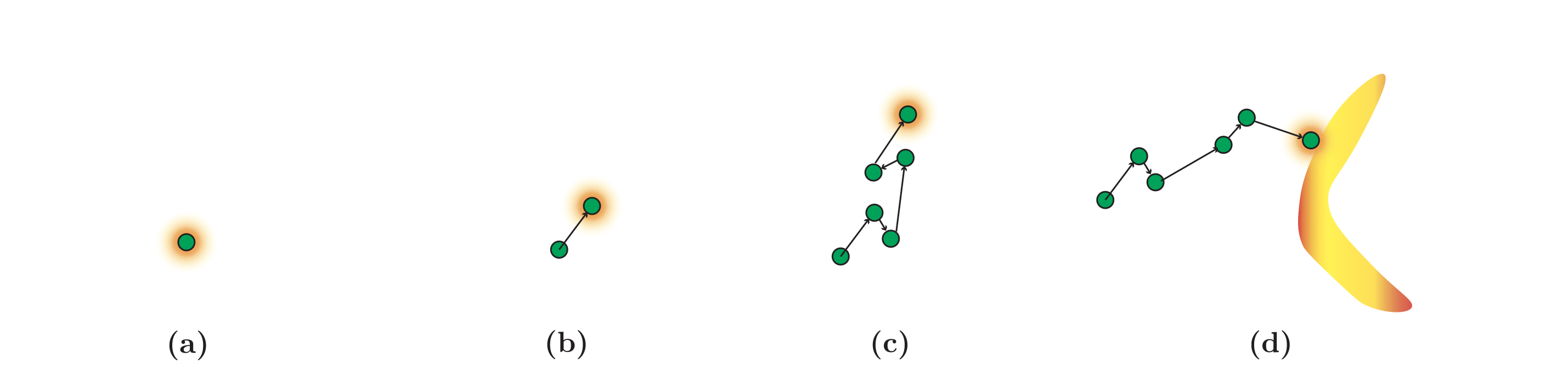}
    \vspace{-20pt}
    \caption{To create a Markov chain, we start with a random sample (a) that is generated with a given proposal (shown in orange). (b) This sample becomes the current state. The next sample (new state) is then generated from the current state following the proposal. (c) As the chain continues to grow, we obtain a sequence of samples that represents a Markov chain. (d) When run long enough, this Markov chain can generate samples that matches the target distribution.}
\end{figure*}
%

\subsection{Stochastic differential equations (SDEs)}

{
\setlength{\columnsep}{4mm}%
\setlength{\intextsep}{4pt}%
\begin{wrapfigure}{r}{0pt}
    \includegraphics[width=0.2\textwidth]{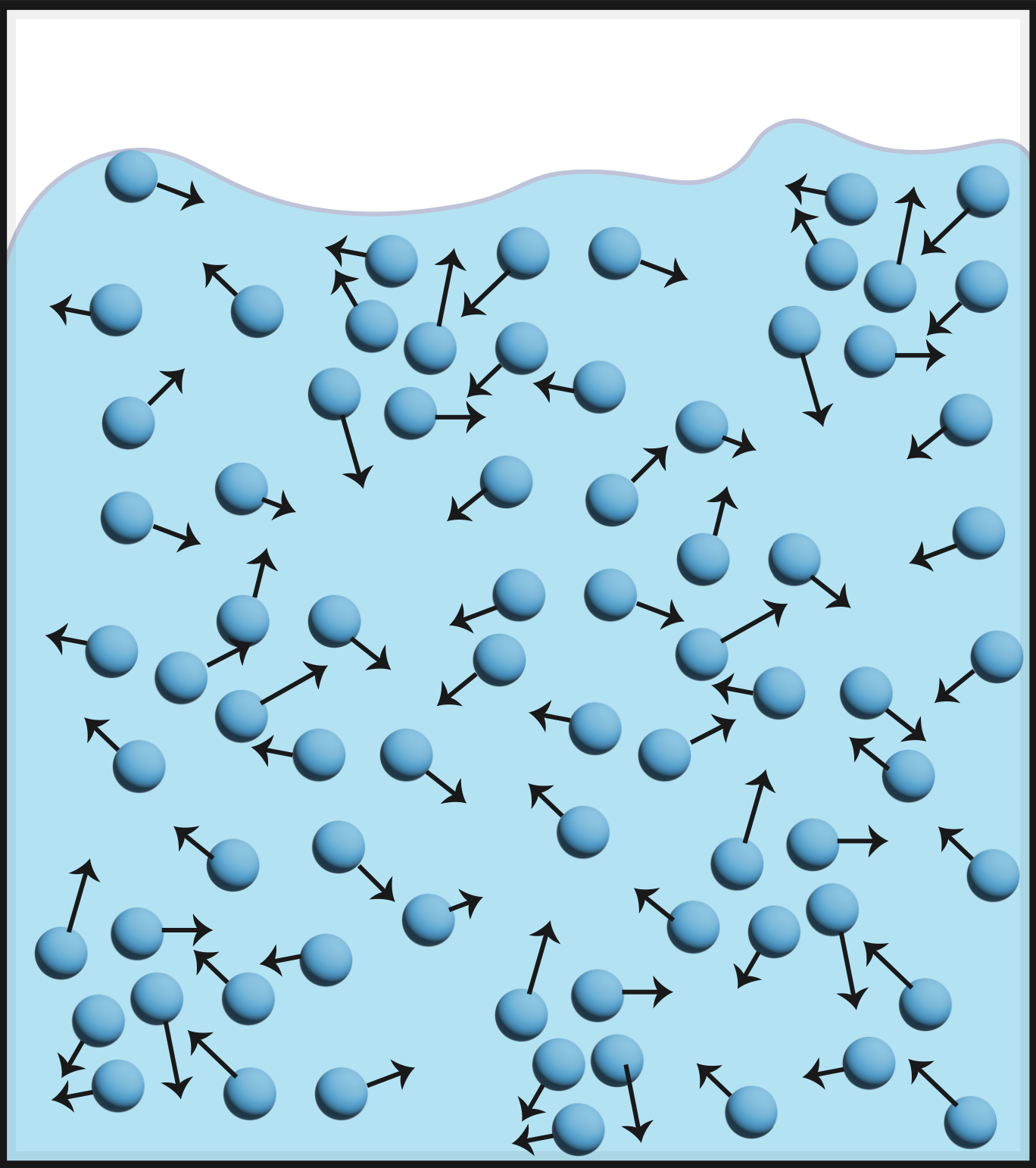}
\end{wrapfigure}
SDEs describe systems that evolve over continuous time with randomness (noise) involved. SDEs often model complex phenomena like stock prices or physical systems influenced by random factors. It describes how we simulate motion numerically. The motion could be of organic "microscopic" particles or of inorganic particles.

Consider particles jiggling in the water (shown on the right). 
There are a huge number of water molecules ($\sim 10^{24}$). 
Each particle is on average moving along certain trajectory (\ie \emph{drift}) with some jiggle~(randomness or \emph{diffusion}). SDEs can be used to simulate such stochastic motion. Even though the particles are jiggling in random directions, in ensemble, their motion can be well-predicted. An SDE typically takes the form~\citep{roberts2002langevin}:
\begin{align}
    d\sample_t = \underbrace{\driftterm(\sample_t,t)}_{\text{drift}} dt + \underbrace{\diffusionterm(\sample_t,t)}_{\text{rate of diffusion}} \!\!\!\!\!\! d\wienerprocess_t
\end{align}
where $\sample_t$ is the state variable at time $t$, $\driftterm(\sample_t,t)$ is the drift term, representing the deterministic part of the dynamics, $\diffusionterm(\sample_t,t)$ is the diffusion term,  representing the stochastic part of the dynamics and $\wienerprocess_t$ is a Wiener process or Brownian motion, which introduces randomness into the system.} 

The solution of an SDE can exhibit the \emph{Markov} property, where future behavior depends only on the present state and not on the path that led to it. In continuous time, the evolution of states in an SDE can be thought of as a continuous Markov process. 

\subsubsection{Brownian motion}

\begin{figure*}[h]
    \centering

\newcommand{\PlotSingleImage}[1]{%
        \begin{scope}
            \clip (0,0) -- (2.5,0) -- (2.5,2.5) -- (0,2.5) -- cycle;
            \path[fill overzoom image=figures/brownian_motion/#1] (0,0) rectangle (2.5cm,2.5cm);
        \end{scope}
        \draw (0,0) -- (2.5,0) -- (2.5,2.5) -- (0,2.5) -- cycle;
}

\newcommand{\scaleval}{1.35}    
\small
\hspace*{-4mm}
\begin{tabular}{c@{\;}c@{\;}c@{\;}c@{\;}c@{}}
\begin{tikzpicture}[scale=\scaleval]
    \PlotSingleImage{bmotion_square.png}
\end{tikzpicture}
&
\begin{tikzpicture}[scale=\scaleval]
    \PlotSingleImage{bmotion_disk.png}
\end{tikzpicture}
&
\begin{tikzpicture}[scale=\scaleval]
    \PlotSingleImage{bmotion_ring.png}
\end{tikzpicture}
&
\begin{tikzpicture}[scale=\scaleval]
    \PlotSingleImage{bmotion_circle.png}
\end{tikzpicture}
&
\begin{tikzpicture}[scale=\scaleval]
    \PlotSingleImage{bmotion_lamp.png}
\end{tikzpicture}
\\
(a) & (b) & (c) & (d) & (e)
\end{tabular}
    \caption{
    Brownian motion for different target densities. Simulating Brownian motion is the same as performing a random walk (a). However, we can restrict the motion along certain trajectories (b-e). The red dot represents the starting point, and the blue dashed line (b,d) or the blue region (c) represents the targetted region/trajectory. The motion is simulated to partially fill the space to demonstrate the evolution of the Markov chain.
    }
    \label{fig:brownian_motion}
\end{figure*}
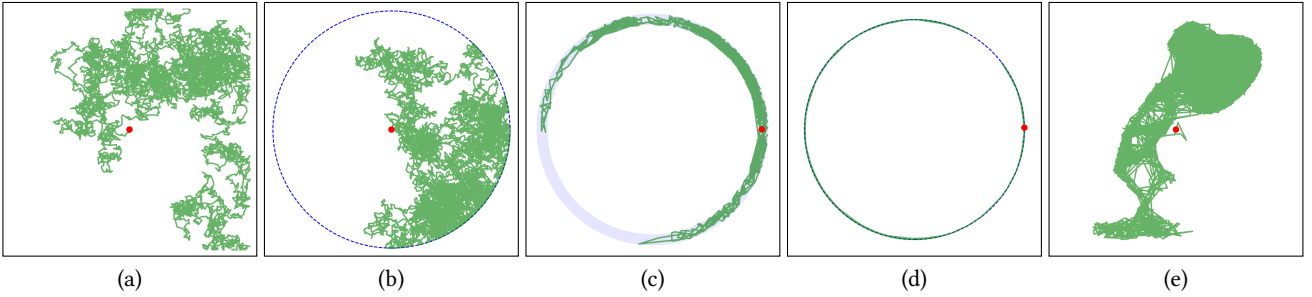

{
\setlength{\columnsep}{4mm}%
\setlength{\intextsep}{4pt}%
\begin{wrapfigure}{r}{0pt}
    \includegraphics[width=0.4\textwidth]{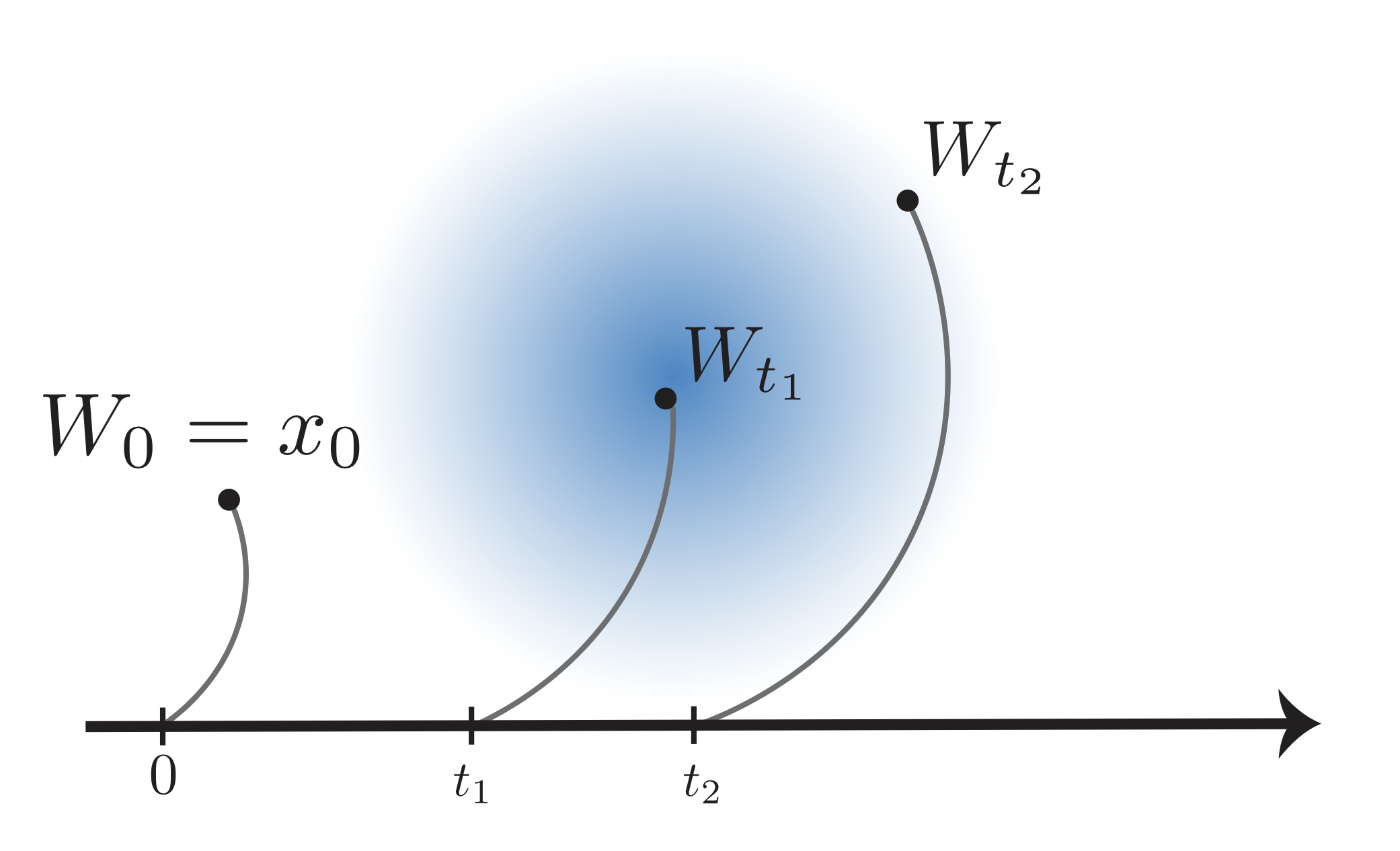}
\end{wrapfigure}
Brownian motion is the simplest form of an SDE and serves as the foundation for more complex models (like Langevin dynamics). It is also known as a Wiener process. It is a continuous-time stochastic process with the following properties:
\begin{itemize}
    \item $\wienerprocess_0 = x_0$.
    \item $\wienerprocess_{t_1} $ has independent increments.
    \item $\wienerprocess_{t_2} - \wienerprocess_{t_1} \sim \normaldistribution(0,t-s)$ for $0 \leq t_1 < t_2$.
    \item $\wienerprocess_t$ varies continuously wrt to $t$.
\end{itemize}
It is a fundamental stochastic process that describes the random motion of particles suspended in a fluid.}
The motion is typically modeled as a stochastic process or random walk.
The motion is completely random with no memory of past states, and it's often modeled using the Fokker-Planck equation for probability density evolution or as a pure diffusion process. 
It is used to describe motion at very small scales (e.g., molecular scales) where thermal fluctuations dominate.
In~\cref{fig:brownian_motion}, we simulate the random walk in space. Although the motion is fully stochastic, it can be directed to follow certain constraints. 
%

\subsubsection{Langevin Dynamics}
Langevin dynamics describes the motion of a particle under the influence of both deterministic forces and random noise. 
It extends Brownian motion by adding a drift term representing deterministic forces and a damping term. It models systems where particles experience both random thermal fluctuations and deterministic forces.
The Langevin equation is an SDE and is given by
\begin{align}
    \label{eq:langevin_equation}
    \lambda \frac{d\sample_t}{dt} = -\frac{d\potentialenergy(\sample)}{d\sample} + \eta(\sample)
\end{align}
where $\potentialenergy(x)$ is the potential energy as a particle's position $\sample_t$ and $\eta(x)$ is the noise term. 
The potential energy term represents the influence of an external or internal force that depends on the particle's position and is typically associated with physical interactions like gravity, electrostatic forces, or molecular bonds. 
The dynamics of the Langevin equation~\cref{eq:langevin_equation} can be written as:
\begin{align}
    \label{eq:langevin_dynamics_potential}
    d\sample_t = - \nabla \potentialenergy(\sample) dt + \sqrt{2} \; d\wienerprocess_t 
\end{align}
where $d\wienerprocess_t$ represents the time derivation of the standard Brownian motion.
The potential energy term guides the particles along the target distribution  $\targetdistribution(x)$. The key relationship between potential energy $\potentialenergy(x)$ and the probability distribution $\targetdistribution(x)$ is given by the Boltzmann distribution in thermal equilibrium $\targetdistribution(x) \propto \exp(-\beta \potentialenergy(x))$. 
The Boltzmann distribution tells us that the probability of a particle being in a particular position is exponentially related to its potential energy at that position. Particles are more likely to be found in regions of lower potential energy.
Therefore, we can set the potential energy to the negative of the logarithm of the target distribution in~\cref{eq:langevin_dynamics_potential}. The resulting Langevin dynamics has the form:
\begin{align}
    \label{eq:langevin_dynamics}
    d\sample_t = \nabla \log \targetdistribution(\sample) + \sqrt{2} \; d\wienerprocess_t 
\end{align}
where $\nabla \log \targetdistribution (\sample_t)$ is the log-probability of the target distribution $\targetdistribution$ and $\wienerprocess_t$ is brownian motion or the Wiener process.

\paragraph{Discretizing SDEs.}
The Euler-Maruyama method is a numerical scheme used to approximate the solution of stochastic differential equations (SDEs). This method is particularly useful in simulating stochastic processes like those described by Langevin dynamics. 
The idea behind discretizing the SDE is to approximate the continuous process $\sample(t)$ at discrete time steps $t_0,t_1,t_2,\cdots$ where $t_{n+1} = t_n + \Delta t$ with time step $\Delta t$. The discretized SDE can have the form:
\begin{align}
    \sample_{n+1} = \sample_n + \driftterm \Delta t + \sigma \sqrt{\Delta t} \xi_n
\end{align}
where $\xi \sim \mathcal{N}(0,1)$ is a normally distributed random number. 
It is a simple and widely used method for solving SDEs, though it may require small time steps for accurate results in systems with strong noise or high variability.

\begin{figure*}
    \centering
     \includegraphics[width=\textwidth]{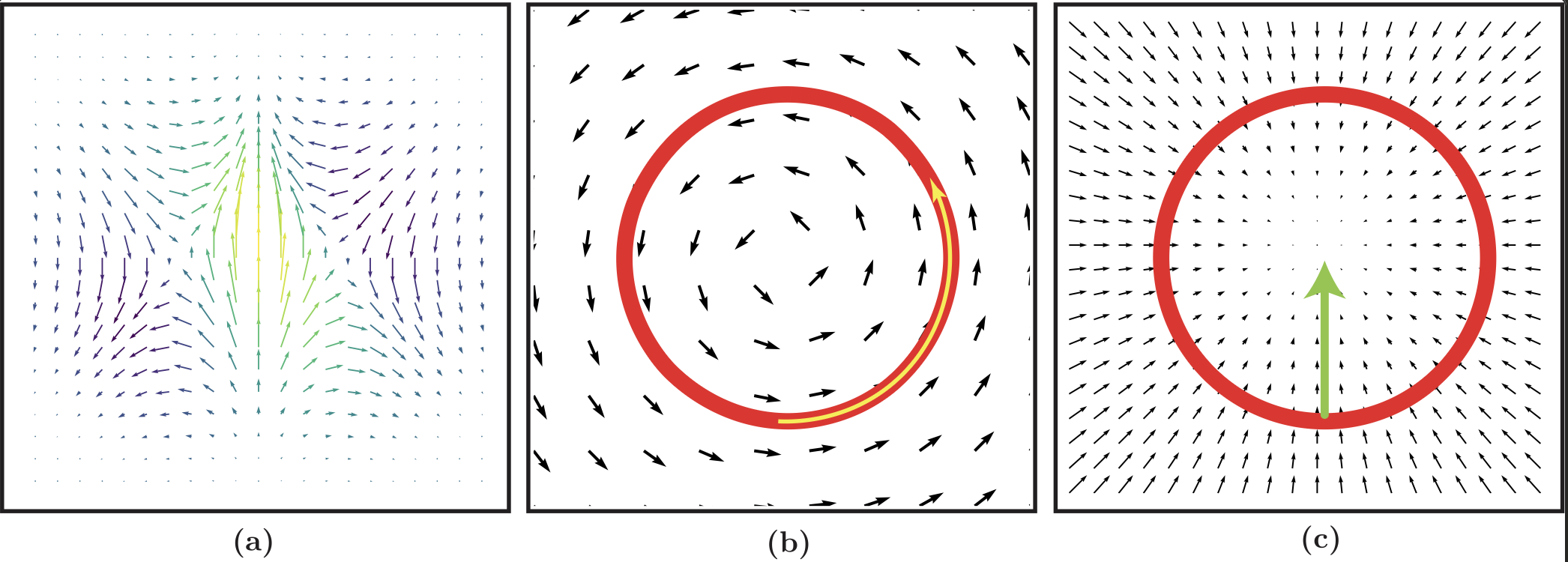}
     \vspace{-20pt}
    \caption{
    Access to the gradients (vector field) of the target distribution (a) is a natural information we can have for exploration. Most Markov transitions are diffusive in nature, \ie the they can spend too much time near the initial point. 
    In order to make large jumps away from the initial point, and into new, unexplored regions of the typical region, we need to exploit information about the geometry. Hamiltonian dynamics is the unique procedure for automatically generating this
coherent exploration for sufficiently well-behaved target distributions. It not only allows efficient movement in the neighborhood of a mode (b) but also towards the mode that needs to be explored (c). The figure is inspired from~\cite{betancourt2018conceptual}.
    }
    \label{fig:langevin_hmc}
\end{figure*}
\subsubsection{Hamiltonian Dynamics}
In Hamiltonian dynamics, we describe a system using a function called the Hamiltonian, which usually represents the total energy (kinetic + potential) of the system. Hamiltonian dynamics uses coordinates (where something is) and momentum (how fast it's moving and in what direction) to describe motion. This shift helps in analyzing complex systems more easily~\citep{betancourt2018conceptual}.

Consider, for example, a typical problem of exploring the target distribution density during optimization. Usually, the information we have is the differential structure of the target distribution which we can query through the \emph{gradient} of the target distribution density. In particular, the gradient defines a vector field in parameter space sensitive to the structure of the target distribution~\cref{fig:langevin_hmc}(a). 
The gradients, however, can only guide us towards the parametrization-sensitive neighborhood like towards the mode (high density region)\cref{fig:langevin_hmc}(c), but not in the parameterization-invariant neighborhoods.
To ensure, we can explore all regions of the distribution we need to ensure a coherent exploration of the geometry of the typical set (the red rings in~\cref{fig:langevin_hmc}b,c). 

Following~\cite{betancourt2018conceptual}, we can think of this exploration as if launching a satellite in a stable orbit around a planet. To ensure the satellite stays within the orbit, we need to give sufficient \emph{momentum} to counteract the gravitational attraction. In probabilistic perspective, this gravitational pull can be thought of as the gradient vector field guiding the exploration. If sufficient momentum is not injected, the exploration will always be moving towards the mode. We can introduce momentum in the probabilistic structure using auxillary momentum parameters. But we have to ensure that the probabilistic structure ensures conservative dynamics. Conservative dynamics in physical systems requires that volumes are exactly preserved. 
Hamiltonian dynamics provides us this procedure to introduce such auxillary momentum.

\paragraph{Phase space.}
Instead of dealing with target parameter space, the state of a Hamiltonian system is represented in a space called phase space, where each point corresponds to a unique combination of position and momentum. The points are \emph{lifted} from the parameter space to the phase space, undergoes trajectory exploration and then projected back to the parameter space.

\paragraph{Hamilton's equation.}
The dynamics of the system are governed by two main equations, known as Hamilton's equations, which tell us how the position and momentum change over time. They can be written as:
\begin{align}
    \frac{d \generalcoord}{dt} = \frac{\partial\hamiltonian(\generalcoord,\generalmomentum)} {\partial\generalcoord} \;,\;
    \frac{d \generalmomentum}{dt} = - \frac{\partial\hamiltonian(\generalcoord,\generalmomentum)}{\partial\generalmomentum}
\end{align}
where $\generalcoord$ is the generalized coordinate (position), $\generalmomentum$ is the generalized momentum and $\hamiltonian(\generalcoord,\generalmomentum)$ is the Hamiltonian, representing the total energy of the system.

\paragraph{Hamiltonian function.}
The Hamiltonian function $\hamiltonian(\generalcoord,\generalmomentum)$ represents the total energy of the system and is typically given by:
\begin{align}
    \hamiltonian(\generalcoord,\generalmomentum) = \potentialenergy(\generalcoord) + \kineticenergy(\generalmomentum)
\end{align}
where $\potentialenergy$ is the potential energy, often related to the negative log-likelihood of the target distribution 
and $\kineticenergy$ is the kinetic energy.
%
When combined with stochastic elements, it leads to methods like Hamiltonian Monte Carlo, which are used for sampling from complex distributions. We will explore Hamiltonian Monte Carlo in the coming sections.

\subsection{Monte Carlo integration}
Estimating high-dimensional integrals \eg 
$\integral = \int_{\domain} f(x) dx$
is usually achieved by numerical integration methods like Monte Carlo.
Monte Carlo estimator of an integral $\integral$ has a form:
\begin{align}
    \label{eq:mce_is}
    \mce = \frac{1}{N} \sum_{i=o}^{N-1} \frac{f(\sample_i)}{\targetdistribution(\sample_i)}
\end{align}
where $f(\cdot)$ is the function to be integrated over the domain $\domain$, $\targetdistribution$ represents the proposal distribution to extract $\samplecount$ samples $\sample_i$ where $i \in {0,\ldots, \samplecount-1}$.
Such estimation is error prone and is visible as noise in rendered images using physically based light transport rendering. 
Several variance reduction strategies are proposed in the literature~\cite{veach1998thesis}. Importance sampling is one such strategy and is known to reduce the variance. 
To perform importance sampling, samples are drawn from a proposal distribution and weighted to be used in the estimator~\cref{eq:mce_is}.
This strategy can be quite efficient if proposal distribution is well-matched with the target distribution. However, finding a right proposal distribution is not always trivial. Also, importance sampling can become impractical for high-dimensional problems where finding a good proposal distribution is challenging. This is where MCMC sampling methods plays a crucial role.

\subsection{MCMC sampling methods}
MCMC method is a powerful tool to generating samples from complex arbitrary distributions. These samples can then be used to approximate integrals such as~\cref{eq:mce_is}.
MCMC is a class of algorithms that generate samples from a target distribution by constructing a Markov chain that has the target distribution as its equilibrium distribution. MCMC algorithm works as follows:
\begin{itemize}[leftmargin=3.4mm]
    \item Initialization: Start with an initial state $\sample_0$
    \item Transition: Define a transition mechanism (often a probability distribution) to move from the current state $\sample_t$ to  a new state $\sample_{t+1}$. Common algorithms include the Metropolis-Hastings algorithm~\citep{veach1997metropolis}.
    \item Stationarity: Ensure that the Markov chain has the target distribution as its stationary distribution. Over time, the distribution of the samples from the chain will converge to the target distribution.
    \item Sampling: Collect samples after a burn-in period (initial samples are discarded to allow the chain to converge).
\end{itemize}
In this part, we will introduce Metropolis-Hastings, Langevin Monte Carlo and Hamilotnian Monte Carlo sampling methods.  
MCMC sampling methods can be used for very complex and high-dimensional distributions where direct sampling is difficult.
Different MCMC algorithms can be tailored for specific types of target distributions. 
We will also highlight their shortcomings \eg MCMC methods require many iterations to converge to the target distribution, which can be computationally expensive.

\subsubsection{Metropolis-Hastings algorithm}
The Metropolis-Hastings algorithm is an MCMC method for obtaining a sequence of samples from a probability distribution from which direct sampling is difficult. 
Here’s a step-by-step algorithm for the Metropolis-Hastings method: 
\begin{itemize}
    \item Initialize the state, 
\item For each iteration, propose a new state from the proposal distribution, 
\item Compute the acceptance ratio and accept or reject the proposed state based on a random draw from a uniform distribution.
\end{itemize}

A python pseudo code is shown here:
%
\begin{lstlisting}[language=Python, caption=Python example]
import numpy as np
import matplotlib.pyplot as plt

# Target distribution: e.g. 2D Gaussian
def target_distribution(x):
    return np.exp(-0.5 * np.dot(x, x))

# Proposal distribution: e.g. Gaussian with mean at the current state
def proposal_distribution(x, sigma=1.0):
    return x + np.random.normal(scale=sigma, size=x.shape)

# Metropolis-Hastings algorithm
def metropolis_hastings(initial_state, num_samples, proposal_sigma):
    samples = []
    x_t = np.array(initial_state)
    
    for _ in range(num_samples):
        x_star = proposal_distribution(x_t, proposal_sigma)
        
        acceptance_ratio = min(1, target_distribution(x_star) / target_distribution(x_t))
        if np.random.rand() < acceptance_ratio:
            x_t = x_star
            samples.append(x_t.copy())
    
    return np.array(samples)

# Parameters
initial_state = [0.0, 0.0]
num_samples = 10000
proposal_sigma = 1.0

# Generate samples using Metropolis-Hastings
samples = metropolis_hastings(initial_state, num_samples, proposal_sigma)
print(samples.shape)

# Plot the samples
plt.figure(figsize=(8, 8))
plt.plot(samples[:, 0], samples[:, 1], 'o', markersize=1)
plt.title('Metropolis-Hastings Samples')
plt.xlabel('x[0]')
plt.ylabel('x[1]')
plt.show()
\end{lstlisting}
%

\subsubsection{Langevin Monte Carlo sampling}
Langevin Monte Carlo (LMC) is a class of Markov Chain Monte Carlo (MCMC) algorithms that generate samples from a probability distribution of interest by simulating the Langevin Equation. It leverages the gradient of the log-probability (log-likelihood) to guide the sampling process more effectively. The update rule for LMC is given by:
\begin{align}
    \label{eq:lmc}
    \sample_{t+1} = \sample_t + \frac{\timestep^2}{2} \nabla \log \targetdistribution (\sample_t) + \timestep \noiseterm_t 
\end{align}
where $\sample_t$ is a current sample, $\timestep$ is the time step, $\nabla \log \targetdistribution (\sample_t)$ is the log-probability of the target distribution $\targetdistribution$ and $\noiseterm_t$ is the noise term, typically drawn from a standard normal distribution $\normaldistribution(0,I)$.

\subsubsection{Annealed Langevin Monte Carlo sampling}
Annealed Langevin Monte Carlo (ALMC) is a variant of LMC that introduces an annealing schedule. Annealing is a process where the "temperature" of the system is gradually decreased, starting from a high value (which allows the sampler to explore the state space more freely) and gradually lowering it to focus more on high-probability regions.

In ALMC, the target distribution is tempered by introducing a temperature parameter \temperature, and this parameter is gradually annealed. The tempered distribution is given by: $\targetdistribution_\temperature(\sample) \propto \targetdistribution(\sample)^{1/\temperature}$. 
The update rule for ALMC is similar to LMC but incorporates the temperature parameter:
\begin{align}
    \label{eq:almc}
    \sample_{t+1} = \sample_t + \frac{\timestep^2}{2\temperature} \nabla \log \targetdistribution (\sample_t) + \timestep \noiseterm_t 
\end{align}
where $\temperature$ decreases over time according to a schedule. Typically, $\temperature$ starts from a high value and gradually decreases to 1.

While LMC is effective for sampling from \emph{unimodal} distributions or when the modes are well-separated and the gradient information is reliable.
On the other hand, ALMC is useful for sampling from \emph{multimodal} distributions where the modes are not easily separable, as the annealing process helps in exploring the global structure before focusing on high-probability regions.


\subsubsection{Hamiltonian Monte Carlo sampling}
A direct connection between SDEs and MCMC is also found in the Hamiltonian Monte Carlo (HMC) method. HMC leverages concepts from physics, particularly Hamiltonian dynamics, which are described by differential equations. The dynamics are typically simulated using numerical methods that solve differential equations, ensuring efficient exploration of the target distribution. 

HMC offers a powerful set of Markov transitions that are capable of performing well over a large class of target distributions. The major challenge in implementing HMC is generating the Hamiltonian trajectories themselves. Formally, integrating along the vector field define by Hamiltonian equations is equivalent to solving ordinary differential equations in phase space. However, most of the ODE solvers can \emph{drift} away from the trajectories accumulating error over time. 

We can, however, use the geometry of the phase space to construct extremely powerful family of numerical solvers known as the use \emph{symplectic integrators}~\citep{leimkuhler2005hamiltonian,hairer2013geometric}.
These integrators are robust to drift and enable high-performance implementaiton of HMC method. 
Symplectice integrators are straightforward to implement in practice. If the probabilistic distribution of the momentum is independent of the position, then we can employ a simple \emph{leapfrog integrator}. 
Given a time discretization, the leapfrog integrator simulates the exact trajectory by precise interleaving of discrete momentum and position updates that ensures
exact volume preservation on phase space. 





%

\subsubsection{Discussion}

In summary, MH is simple and flexible but can be inefficient for complex or high-dimensional problems.
Langevin dynamics improves on MH by using gradient information, but requires smoothness and introduces discretization issues.
HMC is the most efficient in high dimensions, particularly for smooth distributions, but is computationally expensive and requires careful tuning. We provide a detailed comparison among these methods  along different aspects in~\cref{tab:comparing_mcmc_methods}.

MH can be applied to a wide variety of distributions as long as you can evaluate the target distribution's density up to a normalizing constant.
The algorithm is conceptually straightforward and easy to implement.
It can be used with any proposal distribution, allowing for customization to the problem at hand. It guarantees to converge to the true target distribution (if ergodic conditions are met). However, if the proposal distribution is poorly chosen, it may explore the space inefficiently (slow mixing and high autocorrelation). The performance of the algorithm depends heavily on the proposal distribution and its scale. Poor choices can result in high rejection rates. MH can be slow for high-dimensional problems, as proposals often move inefficiently in large spaces.

Langevin dynamics can take advantage of the gradient of the log-posterior, making it more efficient than random-walk methods like MH, especially for smooth target distributions. The gradient information helps proposals move toward high-probability regions, leading to faster exploration of the space. Proposals are more informed, so they generally require less tuning than vanilla MH.
On the other hand, calculating the gradient can be expensive, especially in high-dimensional or complex models.
LMC assumes that the log-posterior is differentiable and smooth. It may not work well with highly non-smooth distributions.
Since Langevin dynamics is a continuous process, discretizing it to use in practice introduces bias unless corrected (e.g., through the Metropolis-adjusted Langevin algorithm, MALA).

Finally, HMC can make large, informed jumps in the parameter space, allowing it to explore high-dimensional spaces much more efficiently than random-walk-based methods like MH.
The proposals are designed to make larger, more informed steps, which reduces autocorrelation and improves convergence speed.
Like Langevin dynamics, HMC leverages gradients to propose new states, making it well-suited for smooth, differentiable posterior distributions. 
However, HMC depends on parameters like step size and the number of leapfrog steps. Poor tuning can lead to either rejection of proposals or inefficient exploration. Each iteration requires solving Hamiltonian dynamics via the leapfrog integrator, which is computationally expensive in high-dimensional models.
HMC can struggle to explore multimodal distributions, as its trajectories are deterministic and may not easily traverse between modes. 

\begin{table}[t]
\centering
\begin{tabular}{|p{23mm}|p{45mm}|p{45mm}|p{45mm}|}
  \hline
  \rowcolor{lightgray}  
  \textbf{Aspect} & \textbf{Metropolis-Hastings (MH)} & \textbf{Langevin Monte Carlo (LMC)} & \textbf{Hamiltonian Monte Carlo (HMC)} \\
  \hline
  Efficiency      & Low for high dimensions (random walk behavior)      & Better than MH, guided by gradients      & Very efficient, especially in high dimensions      \\
  \hline
  Gradient usage      & No      & Yes      & Yes      \\
  \hline
  Autocorrelation      & High, especially with poor proposals      & Lower than MH      & Very low due to long, informed trajectories     \\
  \hline
  Tuning required & Yes (proposal distribution) &	Yes (step size for discretization)	& Yes (step size, number of leapfrog steps) \\
  \hline
  Scalability	& Poor in high dimensions	& Moderate	& Good for large dimensions \\
  \hline
  Applicability	& Very general	& Requires smoothness and gradients	& Requires smoothness and gradients \\
  \hline
  Computational cost per step &	Low	 & Moderate	& High \\
  \hline
  Convergence	& Slower, especially for bad proposals &	Faster, thanks to gradient guidance	& Fast, but sensitive to parameter settings \\
  \hline
  Suitability for multimodality	& Decent, but still slow	& May struggle with multimodal targets	& Poor, deterministic paths struggle with multiple modes \\
  \hline
\end{tabular}
\caption{
\label{tab:comparing_mcmc_methods}
Comparing MH, LMC and HMC sampling methods}
\end{table}

\section{MCMC in rendering}
\label{sec:mcmc_rendering}

Physically based rendering algorithms simulate the behavior of light to turn
scene representations describing object shape and optical properties
into realistic images. For this, they must simulate various physical laws that
can be roughly classified into \emph{transport} and \emph{scattering}, i.e.,
the propagation of light through space, and its local interaction with the
objects comprising the scene. From a mathematical perspective, the entire
problem reduces to evaluating a series of high-dimensional integrals to
determine the radiance $I_j$ received by each pixel $j$ of a virtual camera
observing the scene, i.e.:
\begin{equation}
    \label{eqn:rendering-integral}
    I_j =\int_{\domain}f_j(\vx)\,\mathrm{d}\vx.
\end{equation}
The function $f_j$ characterizes this process for given pixel $j$, while
$\domain$ depends on the specific formulation and algorithm being used. The
dimension of $\domain$ is generally proportional to the number of subsequent
scattering events that should be considered. This number can be rather large
and ranges from tens to many thousands of dimensions to deal with highly
scattering materials like clouds, milk, skin, etc. For this reason, Monte Carlo
methods have become the method of choice in the last decades.

\subsection{Path-space}
\cite{veach1998thesis} introduced a particularly general variant of the
integration problem from Equation~\ref{eqn:rendering-integral} known as the
\emph{path space formulation} that we discuss here for the special case of
scenes containing only surfaces (i.e., lacking volumetric effects).
It decomposes the integration domain $\pathspace$ into union of subspaces:
\begin{gather}
    \pathspace \coloneqq \bigcup_{n=2}^\infty \pathspace_n,\text{ and}\notag\\
\label{eqn:pathspace-def}
\pathspace_n \coloneqq \set{\vx_1\cdots\vx_n \where \vx_1,\ldots,\vx_n \in \surfaces },
\end{gather}
where $\surfaces$ is the set of surfaces within the virtual scene. Elements
$\bar\vx=\vx_1,\ldots\vx_n \in \pathspace_n$
referred to as \emph{paths} denote potential trajectories that light can
take while traveling from the light source towards the virtual sensor.

The pixel intensity $I_j$ is given by
\begin{align}
\label{eqn:pathspace-integral}
I_j &=\int_{\pathspace_2}f(\vx_1\vx_2)\,\mathrm{d}A(\vx_1, \vx_2)\ +\int_{\pathspace_3}f(\vx_1\vx_2\vx_3)\,\mathrm{d}A(\vx_1, \vx_2, \vx_3)+\ldots.
\intertext{%
Because some paths carry more illumination from the light source to the camera
than others, the integrand $f:\pathspace\to\mathbb{R}$ is needed to
quantify their ``light-carrying capacity''; its definition varies based on
the number of input arguments and is given by Equation~\eqref{eqn:contribfun}. The total illumination $I_j$ arriving at the
camera is often written more compactly as an integral of $f$ over the entire path
space, i.e.:}
\label{eqn:pathspace-integral-short}
&\eqqcolon\int_{\pathspace}
f(\bar{\vx})\,\mathrm{d}A(\bar{\vx}).
\end{align} The definition of the
weighting function $f$ consists of a product of terms---one for each
vertex and edge of the path:
\begin{equation}
\label{eqn:contribfun}
f(\vx_1\cdots\vx_n) = L_e(\vx_1\to\vx_{2})
    \left[ \prod_{k=2}^{n-1} G(\vx_{k-1}\leftrightarrow\vx_{k})\,f(\vx_{k-1}\to\vx_{k}\to\vx_{k+1}) \right]
    G(\vx_{n-1}\leftrightarrow\vx_n)\,W_e^{(j)}(\vx_{n-1}\to\vx_{n}).
\end{equation}
The arrows in the above expression symbolize the symmetry of the geometric
terms as well as the flow of light at vertices. $\vx_i\to\vx_{i+1}$ can
also be read as a spatial argument $\vx_i$ followed by a directional argument
$\overrightarrow{\vx_i\vx_{i+1}}$.
Figure~\ref{fig:pathdemo} shows an example light path
and the different weighting terms.
We summarize their meaning below:
\begin{figure}[t]
    \centering
    \includegraphics[width=.7\columnwidth]{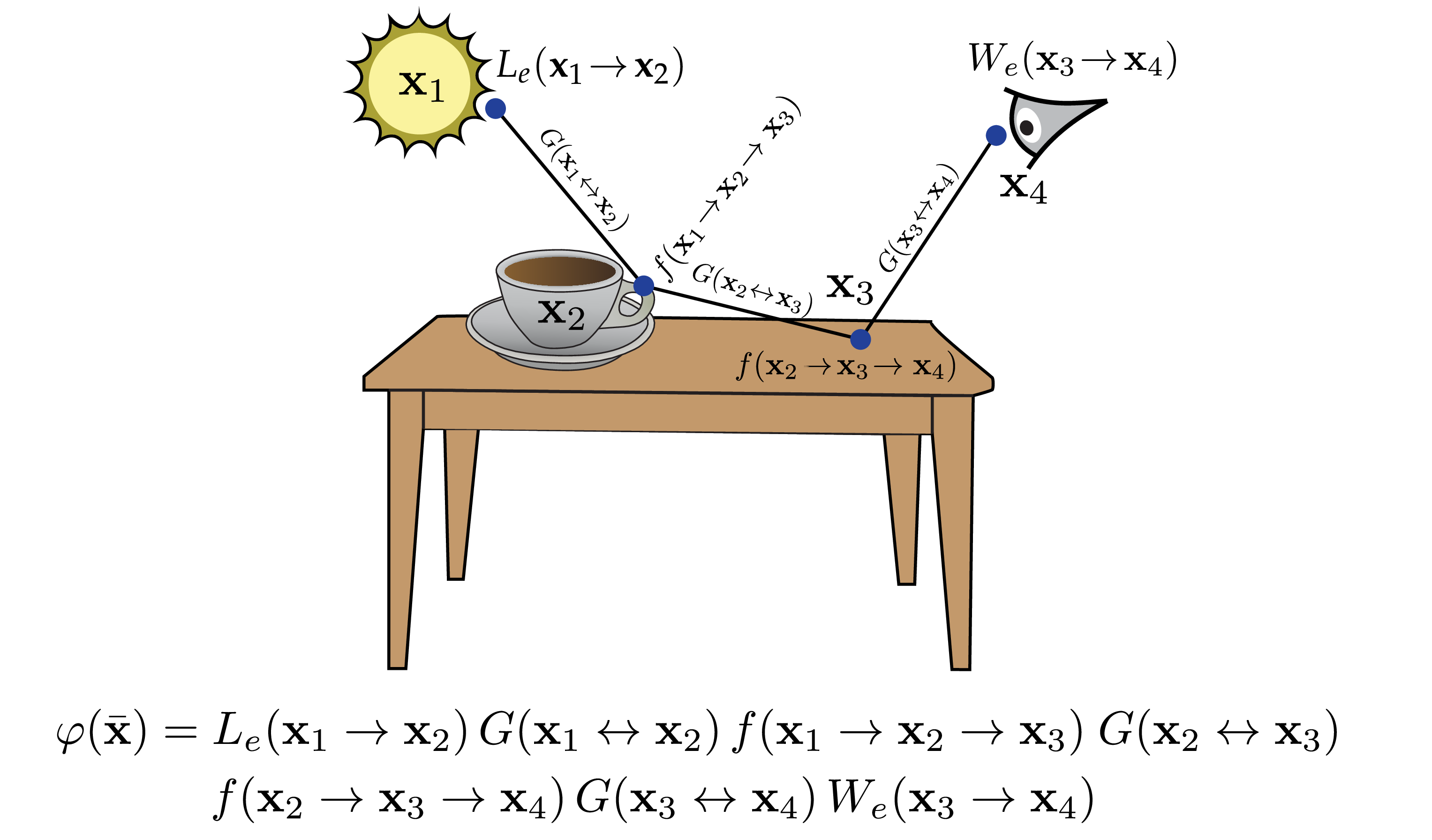}
    \caption{%
        \label{fig:pathdemo}%
        Illustration of a simple light path with four vertices and its
        corresponding weighting function.
    }
\end{figure}
\begin{itemize}
    \item The first term $L_e(\vx_1\to\vx_2)$ is the emission profile of the light source.
        This term models the amount of light emitted from position $\vx_1$
        traveling towards $\vx_2$. It equals zero when $\vx_1$ is not
        located on a light source.

    \item The last term $W_e^{j}(\vx_{n-1}\to\vx_n)$ is the sensitivity profile of pixel $j$ of the camera;
        we can think of the pixel grid as an array of sensors, each with its own profile function.

\item $G(\vx\leftrightarrow\vy)$ is the geometric term, which
        specifies the differential amount of illumination carried along
        segments of the light path. Among other things, it accounts for
        visibility: when there is no unobstructed line of sight between $\vx$
        and $\vy$, $G$ evaluates to zero.

    \item $f(\vx_{k-1}\to\vx_k\to\vx_{k+1})$ is the \emph{bidirectional
        scattering distribution function} (BSDF), which specifies how
        much of the light that travels from $\vx_{k-1}$ to $\vx_{k}$ then
        scatters towards position $\vx_{k+1}$. This function
        characterizes the material appearance of an object (\eg whether it is
        made of wood, plastic, concrete, etc.).
\end{itemize}

Over the last 40 years, considerable research has investigated realistic
expressions for the terms listed above. In this article, we do not
discuss their internals and prefer to think of them as black box functions
that can be queried by the rendering algorithm. This is similar to how
rendering software is implemented in practice: a scene description might
reference a particular material (e.g., car paint) whose corresponding function
$f$ is provided by a library of material implementations.
The algorithm accesses it through a high-level interface shared by all
materials, but without specific knowledge about its internal characteristics.

\subsection{Multiple Importance Sampling}

Many Monte Carlo rendering methods can be interpreted as sampling strategies
that generate paths $\bar\vx$ according to a carefully chosen probability
distribution $p$. Ideally, such a strategy would employ a density function $p$
that is approximately proportional to the integrand $f$, thereby producing
renderings with low variance. However, such strategies are unfortunately not
available in general.

A general building block of sampling strategies are \emph{random walks}:
for example, starting with the endpoint vertex $\vx_{n-1}$ (the position of the
camera), we could randomly sample the predecessor $\vx_{n-2}$ followed by
$\vx_{n-3}$, etc. Similarly, we could generate a random sample on the light
source $\vx_1$ and then work our way towards the camera. Paths could also be
sampled from both sides or the middle (e.g. a window of a room). Combinations
of such walks produce a family of sampling strategies with useful properties.

\wj{Would it be useful to include the pyramid figure showing the different
bidirectional sampling strategies?}
\gurprit{yes, that would help.}
Given a set of sampling strategies on a consistent domain $\domain$, it is
possible to evaluate and compare their densities to combine them effectively.
This is the key insight
of a widely used technique known as \emph{multiple importance sampling}
(MIS)~\citep{Veach:1995:OCS}.

Suppose two statistical estimators of the pixel
intensity $I_j$ are available. These estimators can be used to generate two
light paths $\bar\vx_1$ and $\bar\vx_2$, which have path space probability densities
$p_1(\bar\vx_1)$ and $p_2(\bar\vx_2)$, respectively. The corresponding
MC estimates are given by
\[
    \langle I_j^{(1)}\rangle= \frac{f(\bar\vx_1)}{p_1(\bar\vx_1)}\quad\text{and}\quad
\langle I_j^{(2)}\rangle= \frac{f(\bar\vx_2)}{p_2(\bar\vx_2)}.
\]
To obtain a combined estimator, we could simply average these estimators, i.e.:
\[
    \langle I_j^{(3)}\rangle\coloneqq \frac{1}{2}\big(\langle I_j^{(1)}\rangle + \langle I_j^{(2)}\rangle \big).
\]
However, this is not a good idea, since the combination is affected by the
variance of the worst ingredient estimator. Instead, MIS combines
estimators using weights that are related to the underlying sample density
functions:
\[
    \langle I_j^{(4)}\rangle\coloneqq w_1(\bar\vx_1)\langle I_j^{(1)}\rangle + w_2(\bar\vx_2)\langle I_j^{(2)}\rangle.
\]
A particularly simple weighting function known as the
\emph{balance heuristic} has the following expression for two input strategies:
\begin{equation}
\label{eqn:mis-weights}
    w_i(\bar\vx)\coloneqq
    \frac{p_i(\bar\vx)}{p_1(\bar\vx)+p_2(\bar\vx)}.
\end{equation}
Veach originally showed that no other choice of positive weighting functions can
significantly improve upon the balance heuristic.
\cite{Kondapaneni2019} later introduced a generalization to negative
weights that provably minimizes variance in the general case.

Combinations of multiple sampling techniques are often an effective way to
reduce variance to an acceptable amount. Yet, even such combinations can fail
in simple cases, as we will discuss next.

\subsection{Limitations of Monte Carlo Path Sampling}
\label{sec:limitations-path-sampling}
Ultimately, all Monte Carlo path sampling techniques can be seen to compute
integrals of the weighting function $f$ using a variety of importance
sampling techniques that evaluate $f$ at many randomly chosen points
throughout the domain $\pathspace$.

Certain input, particularly scenes containing metal, glass, or other shiny
surfaces, can lead to integrals that are difficult to evaluate. Depending on
the roughness of the surfaces, the integrand can take on large values over
small regions of the integration domain. Surfaces of lower roughness lead to
smaller and higher-valued regions, which eventually collapse to
lower-dimensional sets with singular integrands as the surface roughness tends
to zero. This case where certain paths cannot be sampled at all is known as the
\emph{problem of insufficient techniques}~ \cite{kollig2002efficient}.

\begin{figure*}[t]
    \centering
    \includegraphics[width=\textwidth]{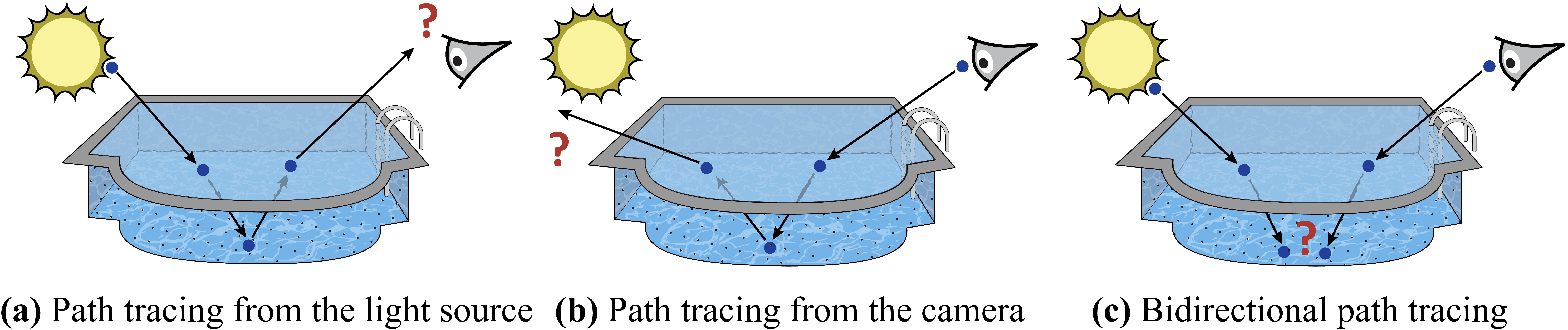}
    \caption{%
        \label{fig:unbiased-difficulties}%
        Illustration of the difficulties of sequential path sampling
        methods when rendering caustic patterns at the bottom of a swimming
        pool. \textbf{(a,\ b)}: Unidirectional techniques sample
        light paths by executing a random walk consisting of alternating
        transport and scattering steps. The only way to successfully complete a
        path in this manner is to randomly ``hit'' the light source or camera,
        which happens with exceedingly low probability. \textbf{(c)}:
        Bidirectional techniques trace paths from both sides, but in this case
        they cannot create a common vertex at the bottom of the pool to join
        the partial light paths.
    }
\end{figure*}

Convergence
problems arise whenever high-valued regions receive too few samples. Depending
on the method used, this manifests as objectionable noise or other visual
artifacts in the output image that gradually disappear as the sample count $N$
tends to infinity. However, due to the slow convergence rate of MC integration
(typical error is $\mathcal{O}(N^{-0.5})$), it may not be an option to wait for
the error to average out. Such situations can force users of rendering software
to make unrealistic scene modifications (\eg disabling certain light
interactions), compromising realism in exchange for obtaining
converged-looking results within a reasonable time.

Figure~\ref{fig:unbiased-difficulties} illustrates the behavior of several
path sampling methods when rendering \emph{caustics}
at the bottom of the swimming pool. This refers to the
light patterns resulting from
focused refraction by ripples in the water surface.

In Figure~\ref{fig:unbiased-difficulties} (a), light tracing samples paths
starting from the light source. This eventually leads to a path segment that
leaves the pool, but it never hits the camera aperture and thus cannot
contribute to the output image. Path tracing in
Figure~\ref{fig:unbiased-difficulties}~(b) generates paths from the opposite
end and also remains extremely inefficient. Assuming for simplicity that rays
leave the pool with a uniform distribution in
the probability of hitting the sun
with an angular diameter of $\sim 0.5^\circ$ is on the order of $10^{-5}$.
Bidirectional sampling methods (Figure~\ref{fig:unbiased-difficulties}~(c))
tracing from both sides also fail: they generate two vertices at the bottom of
the pool as shown in the figure, but these cannot be connected: the resulting
edge would be fully contained in a surface rather than representing transport
\emph{between} surfaces.

The main difficulty in scenes like this is that caustic paths are tightly
constrained: they must start on the light source, end on the aperture, and
satisfy Snell's law in two places. Sequential sampling approaches are able to
satisfy all but one constraint and run into issues when there is no way to
complete the majority of paths.

Paths like the one examined in Figure~\ref{fig:unbiased-difficulties} lead to
poor convergence in other settings as well; they are collectively referred to
as \emph{specular--diffuse--specular} (\texttt{SDS}) paths due to the
occurrence of this sequence of interactions in their path classification.
\texttt{SDS} paths occur in common situations such as a tabletop seen through a
drinking glass standing on it, a bottle containing shampoo or other translucent
liquid, a shop window viewed and illuminated from outside, as well as
scattering inside the eye of a virtual character. Even in scenes where these
paths do not cause dramatic effects, their presence can lead to excessively
slow convergence in rendering algorithms that attempt to account for all
transport paths. It is important to note that while the \texttt{SDS} class of
paths is a well-studied example case, other classes (\eg involving glossy
interactions) can lead to many similar issues. It is desirable that rendering
methods are robust to such situations.

Correlated path sampling techniques based on MCMC offer an attractive way to
approach such challenges because they provide a framework in which the costly
discovery of an SDS path can be amortized by exploring its neighborhood.

\subsection{Metropolis Light Transport}

In 1997, Veach and Guibas proposed an unusual rendering technique named
Metropolis Light Transport \citep{veach1997metropolis}, which applies the
Metropolis-Hastings algorithm to Equation~\ref{eqn:pathspace-integral-short}.
Using correlated samples and highly specialized mutation rules, their approach
enables more systematic exploration of the integration domain, avoiding many of
the problems encountered by methods based on standard Monte Carlo and
sequential path sampling.

MCMC rendering methods in general sample light paths proportional to the amount
they contribute to the pixels of the final rendering; by increasing the pixel
brightness in this way during each iteration, these methods effectively compute
a 2D histogram of the marginal distribution of $f$ over pixel coordinates. This
is exactly the image to be rendered up to a global scale factor, which can be
recovered using a traditional MC sampling technique. The main difference among
these algorithms is the underlying state space, as well as the employed set of
mutation rules.

\begin{figure}[t]
  \centering
  \includegraphics[width=.7\textwidth]{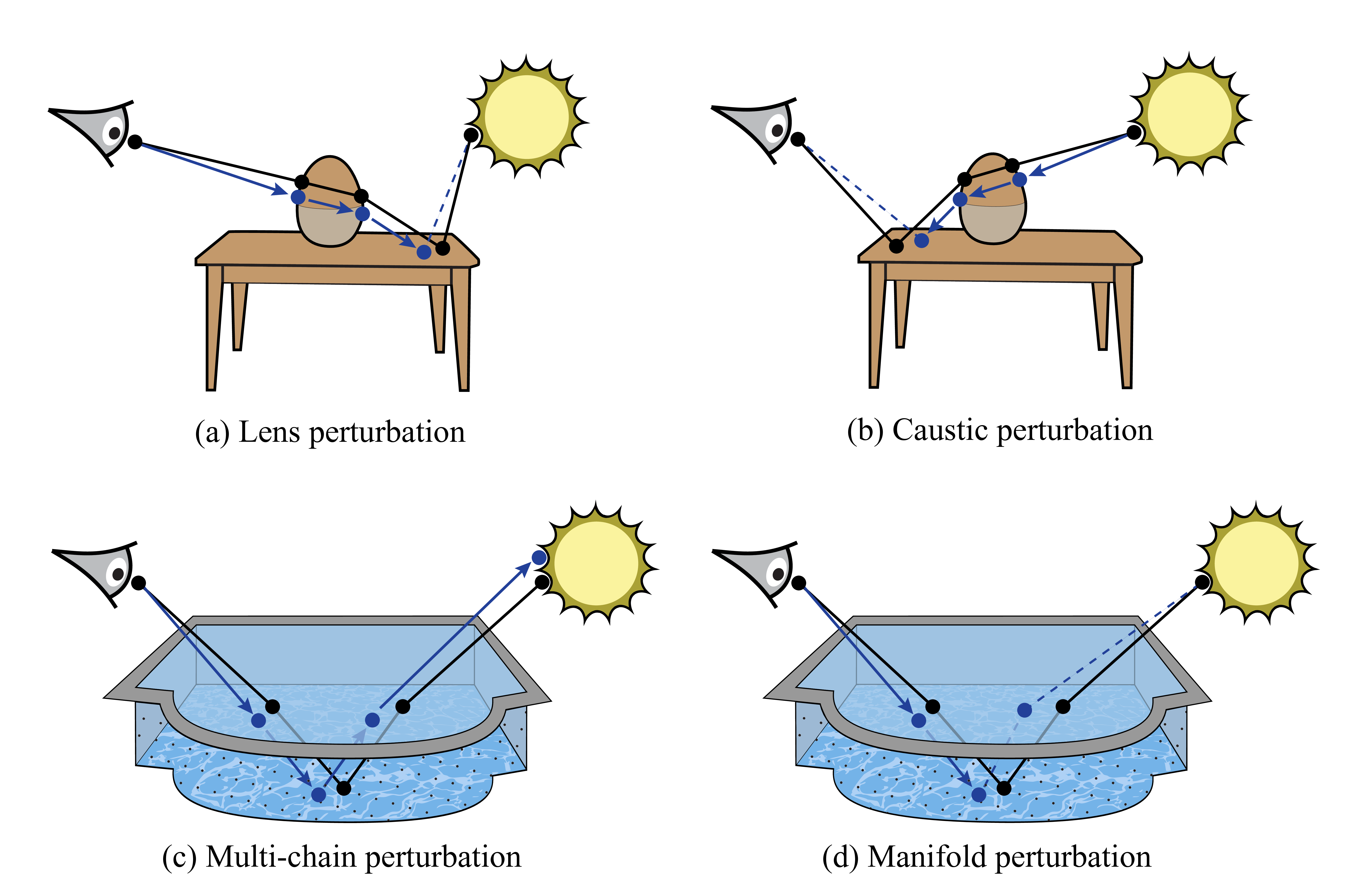}
  \caption{\label{fig:mlt-perturbations}%
    MLT operates on top of path space, which permits the use of a variety of
    mutation rules that are motivated by important physical scattering effects.
    The top row illustrates ones that are useful when rendering a scene
    involving a glass object on top of a diffuse table. The bottom row
    is the swimming pool example from Figure~\ref{fig:unbiased-difficulties}. In
    each example, the original path is black, and the proposal is highlighted
    in blue.
  }
\end{figure}
MLT distinguishes between \emph{mutations} that
change the structure of the path and \emph{perturbations} that move the
vertices by small distances while preserving the path structure, both using the
building blocks of bidirectional path tracing to sample paths. One of the
following operations is randomly selected in each iteration:

\begin{enumerate}
\item \textbf{Bidirectional mutation}:
This mutation replaces a segment of an existing path with a new
segment (possibly of different length) generated by a random walk strategy.
This rule generally has a low acceptance ratio but it is essential to guarantee
ergodicity of the resulting Markov Chain.

\item \textbf{Lens subpath mutation}: The lens subpath mutation is similar to the previous
    mutation but only replaces the \emph{lens subpath}, which is defined as
    the trailing portion of the light path matching the regular expression \texttt{[\textasciicircum S]S*E}.

\item \textbf{Lens perturbation}: This transition rule shown in
    Figure~\ref{fig:mlt-perturbations}a only perturbs the lens subpath rather than
    regenerating it from scratch. In the example, it slightly rotates the
    outgoing ray at the camera and propagates it until the first non-specular
    material is encountered. It then attempts to create a connection (dashed
    line) to the unchanged remainder of the path.
\item\textbf{Caustic perturbation}: The caustic perturbation (Figure~\ref{fig:mlt-perturbations}b) works just like
    the lens perturbation, except that it proceeds in reverse starting at the light source.
    It is well-suited for rendering caustics that are
    directly observed by the camera.
\item\textbf{Multi-chain perturbation}: This transition rule
    (Figure~\ref{fig:mlt-perturbations}c) is used when there are multiple separated
    specular interactions, \eg in the swimming pool example
    encountered before. After an initial lens perturbation, a cascade of
    additional perturbations follows until a connection to the remainder
    of the path can finally be established.
\end{enumerate}
The main downside of MLT is the severe effort needed to implement this method:
several of the mutation and perturbation rules (including their associated
proposal densities) are challenging to reproduce. Another problem is that a
wide range of different light paths generally contribute to the output image.
The MLT perturbations are designed to deal with specific types of light paths,
but it can be difficult to foresee every kind in order to craft a suitable set
of perturbation rules. In practice, the included set is insufficient.

\subsection{Primary sample space MLT}
While path space is a powerful foundation for reasoning about light transport,
the resulting algorithms are notoriously difficult to implement.
\cite{Kelemen:2002:ASA} later showed that a much simpler approach
can be used to directly combine MCMC sampling with existing MC rendering
algorithms, making it possible to altogether avoid the complexities of path
space. The main idea underlying their method named \emph{Primary Sample Space
MLT} (PSSMLT) is general and can also be applied to integration problems
outside of rendering.

PSSMLT operates on top of an existing MC sampling technique that draws
light paths $\bar \vx$ from a target distribution.
At the implementation level, such sampling techniques are usually implemented
by warping uniformly distributed input samples $\vu$ into the desired distribution. All
randomness is provided by the input sample $\vu$, while the warping steps are
not only deterministic but usually even differentiable. In practice, the
inverse transform method is almost always used to realize this warping step.

The rendering process can then be seen compute an integral of a function with
with signature $\Psi:[0,1]^{n}\to\mathbb{R}$ that maps a
vector of univariate samples to a pixel intensity estimate.
By taking many estimates and averaging them to obtain a converged pixel intensity,
path tracing is effectively integrating the estimator
over a $n$-dimensional unit hypercube of ``random numbers'' denoted as \emph{primary sample space}:
\begin{equation}
    \label{eqn:integral-kelemen-mlt}
    I_j=\int_{[0,1]^n}\Psi(\mathbf{\vu})\,\mathrm{d}\mathbf{\vu}.
\end{equation}

\begin{figure}[t]
  \centering
  \includegraphics[width=.7\textwidth]{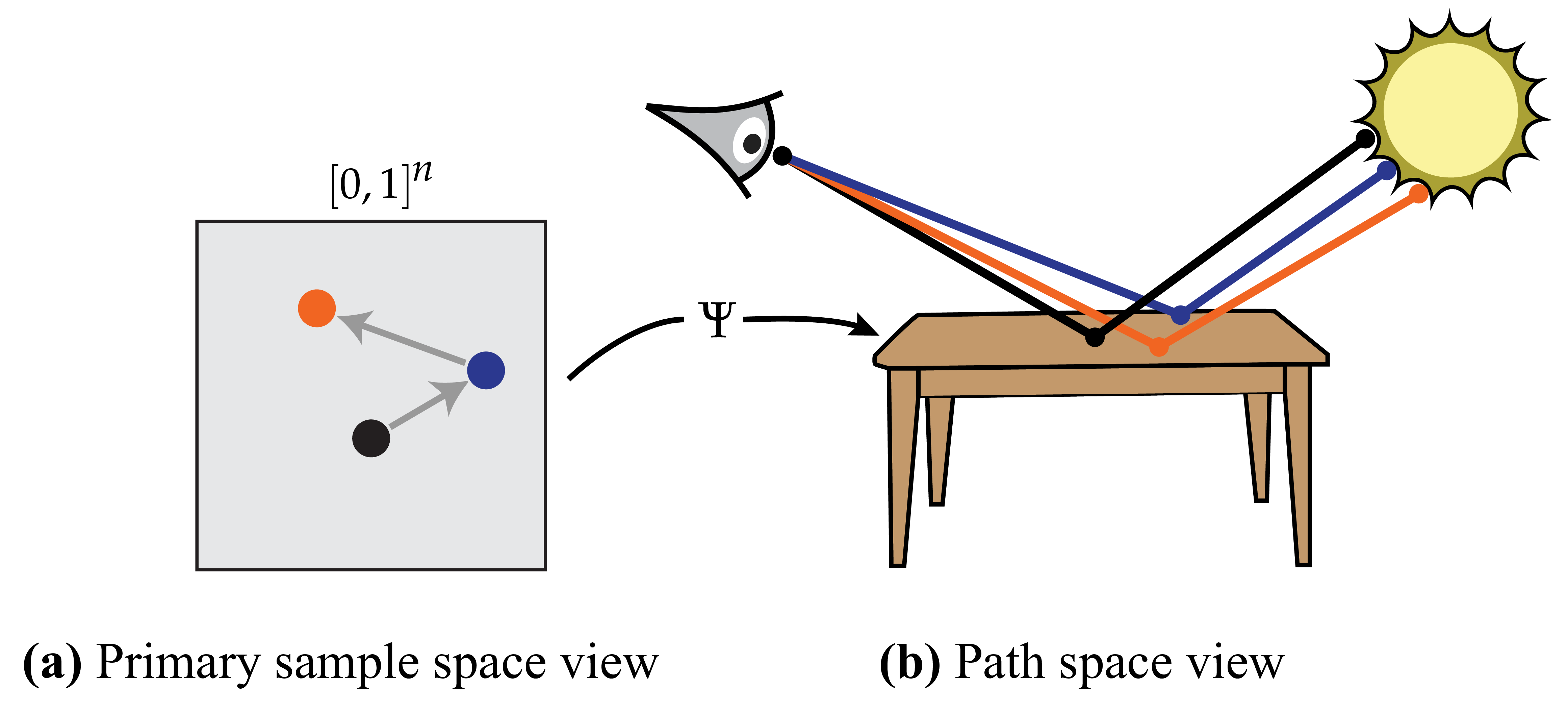}
  \caption{\label{fig:pssmlt-perturbations}%
    Primary Sample Space MLT performs mutations in an abstract random number
    space. A deterministic mapping $\Psi$ induces corresponding
    mutations in path space.
  }
\end{figure}

The key idea of PSSMLT is to compute Equation~\eqref{eqn:integral-kelemen-mlt}
using MCMC integration on directly primary sample space, which leads to a trivial
implementation, as all complications involving light paths and other
rendering-specific details are encapsulated in the ``black box'' mapping $\Psi$ (Figure~\ref{fig:pssmlt-perturbations}).

PSSMLT uses two types of mutation functions. The first is an
independence sampler, i.e., it forgets the current state and switches to a new
set of pseudorandom variates. This is needed to ensure that the Markov Chain is
ergodic. The second is a local (e.g. Gaussian or similar) proposal centered
around a current state $\vu_i\in[0,1]^n$. Both are symmetric so that the
proposal density cancels in the Metropolis-Hastings acceptance ratio.

PSSMLT uses independent proposals to find important light paths that cannot be
reached using local proposals. When it finds one, local proposals are used to
explore neighboring light paths which amortizes the cost of the search. This
can significantly improve convergence in many challenging situations and is an
important advantage of MCMC methods in general when compared to MC integration.

Another advantage of PSSMLT is that it explores light paths through a black box
mapping $\Psi$ that already makes internal use of sophisticated importance
sampling techniques for light paths, which in turn leads to an easier
integration problem in primary sample space. The main disadvantage of this
method is that its interaction with $\Psi$ is limited to a stream of
pseudorandom numbers. It has no direct knowledge of the generated light paths,
which prevents the design of more efficient mutation rules based on the
underlying physics.

\subsection{Hessian-Hamiltonian MLT}
Light paths are often highly constrained because they must satisfy physical
constraints at multiple vertices (Section~\ref{sec:limitations-path-sampling}).
A consequence of such constraints is that the effective exploration of the
neighborhood of a light path requires a concerted change to multiple
coordinates at once. Simple isotropic transition densities that do not
explicitly model dependencies between coordinates tend to be rejected following
steps into low-valued regions of the integrand.

Later work on MCMC rendering algorithms introduced \emph{anisotropic} proposals
to alleviate these problems. \cite{li2015anistropic} considered the behavior of
Hamiltonian dynamics in a neighborhood of the integrand. When the integrand $f$
is locally approximated using a two-term Taylor expansion, a windowed HMC step
reduces to a sample from a multivariate normal distribution, whose covariance
can be computed from the Hessian matrix. The final method resembles MALA with
an anisotropic proposal density.


\section{MCMC in optimization}
\label{sec:mcmc_optimization}

Stochastic gradient descent (SGD) methods are primarily used to optimize functions, particularly in training machine learning models where the objective is to minimize a loss function. 
SGD shares similarities with MCMC sampling methods in their iterative nature, especially in the context of Bayesian inference and machine learning~\citep{chen2016bridging}. 
In this part, we will lay down the similarites between both SGD and MCMC methods and review SGD algorithms driven by MCMC methods.

%

%

\subsection{Bayesian inference}

Bayesian inference is a method of statistical inference in which Bayes' theorem is used to update the probability of a hypothesis as more evidence or data becomes available. It provides a principled way to combine prior knowledge (beliefs about parameters) with new data to make probabilistic conclusions about uncertain parameters or models. The foundations of Bayesian inference is Bayes' theorem, which is expressed as:
\begin{align}
    \targetdistribution(\decoderparams|\text{data}) = \frac{\targetdistribution(\text{data} | \decoderparams) \targetdistribution(\decoderparams)}{\targetdistribution(\text{data})}
\end{align}
where
\begin{itemize}
    \item $\targetdistribution(\decoderparams|\text{data})$ is the \textbf{posterior probability} of the parameter $\decoderparams$ given the observed data. This is what we want to infer. 
    The posterior distribution combines the prior and likelihood to give an updated probability distribution of the parameter $\decoderparams$ after observing the data.
    
    \item $\targetdistribution(\text{data} | \decoderparams)$ is the \textbf{likelihood}, which represents how likely the observed data is under a given parameter $\decoderparams$. The likelihood represents the probability of observing the data given a particular value of $\decoderparams$. It is derived from the model and the observed data
    
    \item $\targetdistribution(\decoderparams)$ is the \textbf{prior} probability, which reflects the prior belief about the parameter $\decoderparams$ before seeing the data. It allows you to incorporate existing knowledge or assumptions about the parameter.
    
    \item $\targetdistribution(\text{data})$ is the \textbf{evidence} (or marginal likelihood), which normalizes the posterior and ensures that the probabilities sum to $1$. The evidence term is the probability of observing the data under all possible values of $\decoderparams$. It is often treated as a normalizing constant and doesn't affect the inference about $\decoderparams$.
\end{itemize}
Bayesian inference provides a powerful framework for updating beliefs based on new data, handling uncertainty, and making probabilistic predictions. By leveraging both prior knowledge and data, it offers a flexible and rigorous approach to statistical inference.

\subsection{Stochastic gradient descent (SGD)}

SGD is an optimization algorithm that iteratively updates model parameters to minimize a loss function. It uses the gradient of the loss function with respect to the model parameters to guide the updates. In each iteration, a mini-batch of data is used to compute an approximate gradient, making the process stochastic. The update rule for SGD is:
\begin{align}
    \label{eq:sgd}
    \param_{t+1} = \param_t - \learningrate \nabla_\param \loss(\param_t) 
\end{align}
where $\param_t$ are the model parameters at iteration $t$,
$\learningrate$ is the learning rate, 
$\nabla_\param \loss(\param_t)$ is the gradient of the loss function $\loss$ with respect to $\param_t$.

SGD is primarily used for deterministic optimization, It is efficient for large-scale machine learning problems due to its stochastic nature. However, it is prone to getting stuck in local minima or saddle points.

\subsection{Stochastic Gradient Langevin Dynamics (SGLD)}

SGLD is an algorithm that combines elements of SGD with Langevin dynamics~\citep{welling2011sgld}, introducing a noise term to the parameter updates. This addition allows SGLD to perform both optimization and sampling from the posterior distribution in Bayesian inference. The update rule for SGLD is very similar to the SGD update rule from~\cref{eq:sgd} and is given by:
\begin{align}
    \label{eq:sgld}
    \param_{t+1} = \param_t - \frac{\learningrate}{2} \nabla_\param \loss(\param_t) + \sqrt{\learningrate} \noiseterm_t
\end{align}
where $\param_t$ are the model parameters at iteration $t$,
$\learningrate$ is the learning rate, 
$\nabla_\param \loss(\param_t)$ is the gradient of the loss function $\loss$ with respect to $\param_t$ and $\noiseterm_t$ is the noise term, typically drawn from a standard normal distribution $\normaldistribution(0,I)$.

\begin{lstlisting}[language=Python, caption=Python example]
# Code snippet for SGLD update rule
import numpy as np

# Loss function and its gradient
def loss_function(theta):
    return 0.5 * np.sum(theta**2)

def gradient_loss_function(theta):
    return theta

# SGLD update rule
def sgld_update(theta, eta):
    grad = gradient_loss_function(theta)
    noise = np.random.normal(size=theta.shape)
    theta = theta - 0.5 * eta * grad + np.sqrt(eta) * noise
    return theta

# Parameters
theta = np.array([2.0, -3.0])  # Initial parameters
eta = 0.01  # Learning rate
num_iterations = 1000

# Perform SGLD
samples = []
for _ in range(num_iterations):
    theta = sgld_update(theta, eta)
    samples.append(theta.copy())

# Convert samples to numpy array for analysis
samples = np.array(samples)

# Plot the samples
import matplotlib.pyplot as plt

plt.plot(samples[:, 0], samples[:, 1], 'o-', markersize=2)
plt.title('SGLD Samples')
plt.xlabel('Theta[0]')
plt.ylabel('Theta[1]')
plt.show()
\end{lstlisting}


\paragraph{SGD vs. SGLD.} There are some key differences: 
\begin{itemize}
    \item Objective: SGD aims to find a point estimate that minimizes the loss function, whereas, SGLD aims to sample from the posterior distribution of the model parameters, making it suitable for Bayesian inference.
    \item Update rule: SGD updates parameters using the gradient of the loss function with respect to the parameters. SGLD, on the other hand, Updates parameters using the gradient of the loss function and adds a noise term to ensure stochasticity.
    \item Noise term: SGD does not include an explicit noise term in the update rule. SGLD includes a noise term, which helps in exploring the parameter space more thoroughly and escaping local minima.
    \item Applications: SGD is used for deterministic optimization tasks, such as training neural networks, whereas, SGLD is used for probabilistic modeling and Bayesian inference, where sampling from the posterior distribution is required.
\end{itemize}

\subsection{Bayesian inference using SGD}

Performing Bayesian inference using Stochastic Gradient Descent (SGD) combines the ideas from Bayesian statistics with optimization techniques like SGD. The general idea is to approximate the posterior distribution over model parameters using gradient-based methods. This approach is particularly useful when exact Bayesian inference (like through Markov Chain Monte Carlo) is computationally infeasible for large datasets or complex models. 

\paragraph{Maximum A Posteriori (MAP) Estimation using SGD}
MAP estimation is a point estimate method in Bayesian inference. Instead of finding the full posterior distribution, we find the parameter value that maximizes the posterior distribution. The MAP objective is:
\begin{align}
    \param_{MAP} = \arg \max_\param \targetdistribution(\param|\text{data}) = \arg \max_\param \targetdistribution(\text{data} | \param) \targetdistribution(\param)
\end{align}
Taking the logarithm of the posterior (since log is a monotonic function), this becomes:
\begin{align}
    \param_{MAP} = \arg \max_\param (\log \targetdistribution(\text{data} | \param) + \log \targetdistribution(\param))
\end{align}
Here, SGD can be used to optimize the MAP objective by computing gradients with respect to $\param$, where $\log \targetdistribution(\text{data})$ is the likelihood (often minimized using SGD), and $\targetdistribution(\param)$ is the prior (often treated as a regularization term, like $\mathcal{L}_2$ regularization).

To implement this optimization, we start with an initial guess for $\param$. 
Then we use SGD~\cref{eq:sgd} to update the parameters by following the gradient of the posterior:
\begin{align}
    \param_{t+1} = \param_t - \learningrate \nabla_\param (-\log \targetdistribution(\text{data} | \param) - \log \targetdistribution(\param))
\end{align}
where $\learningrate$ is the learning rate. 

\paragraph{Using SGLD.}

SGLD adds noise to the gradient updates from SGD to simulate Langevin dynamics, which helps approximate the posterior distribution instead of just finding a point estimate. The update rule for SGLD from~\cref{eq:sgld} is:
\begin{align}
     \param_{t+1} = \param_t - \learningrate \nabla_\param (-\log \targetdistribution(\text{data} | \param) - \log \targetdistribution(\param)) + \sqrt{\learningrate} \noiseterm_t
\end{align}
where $\noiseterm_t$ is a Gaussian noise with zero mean and unit variance. 
The injected noise ensures that the parameter updates sample from the posterior distribution rather than simply converging to a single point (as in MAP estimation). Over time, this stochastic process approximates the posterior distribution.
In the next section, we look at Bayesian (variational) inference using neural networks. In this method, you approximate the posterior distribution with a simpler distribution (often Gaussian) and minimize the divergence between the true posterior and the approximate one.
%

\section{MCMC in generative modeling}

Diffusion models~\citep{sohldickstein2015deep} are the very backbone of modern generative AI pipelines and they are built on the very foundations of MCMC methods. 
In this part, we start by showing how MCMC can be seen as a primitive generative model. We briefly introduce evidence lower bound (ELBO), variational autoencoders (VAEs) and Hierarchical variational autoencoders (HVAEs) that lays the foundations for the variational diffusion models. We then introduce energy-based models which are known to be very flexible 
We then introduce score-based diffusion models that are driven by the SDEs~\citep{song2021score}. 
During this part, we will walk through the skeleton code that will result in a fully functional diffusion model for visual content creation. 


\begin{table}
    \centering
    \small
    \caption{
        Commonly used notations throughout the document.
    }
    \label{tab:Notation}
    \setlength{\tabcolsep}{2.5pt}
    \fontsize{6.9pt}{7pt}
    \begin{tabularx}{\columnwidth}{l X}
        \toprule
        \small \textbf{Notation} & \small\textbf{Description} \\
        \midrule
        $\sample, \latentvar$ & input (observed) data, latent space variables  \\
        $\datadistribution(\sample)$ & distribution of the observed data or likelihood of all observed $\sample$ or the target distribution \\
        $\log\datadistribution(\sample)$ & evidence or log-likelihood of the data \\
        $\encodertrue(\latentvar | \sample)$ & ground truth posterior (encoder) that defines the distribution of latent variables $\latentvar$ over observed samples $\sample$ \\
        $\encoder(\latentvar | \sample)$ & variational posterior (encoder) distribution with parameters $\encoderparams$ that we seek to optimize to match the ground truth $\encodertrue(\latentvar | \sample)$ \\
        $\decoder(\sample | \latentvar)$ & decoder distribution parameterized by learnable parameters $\decoderparams$ \\
        \bottomrule
    \end{tabularx}
    \vspace{-1mm}
\end{table}

\subsection{From variatonal autoencoders to variational diffusion models}
The keyword \emph{variational} is used in mathematical analysis when we deal with maximizing or minizming functionals. Our focus is \emph{likelihood-based} generative models where the idea is to learn a model that maximizes the \emph{likelihood} $\datadistribution(\sample)$ of all observed data $\sample$. We can imagine a joint distribution $\datadistribution(\sample,\latentvar)$ that models the joint probability of the observed samples and their latent variables. There are two ways we can manipulate this joint distribution to recover the likelihood of the observed data distribution $\datadistribution(\sample)$; we can explicitly marginalize the latent variable $\latentvar$:
\begin{align}
    \datadistribution(\sample) = \int \datadistribution(\sample,\latentvar) d\latentvar
\end{align}
which requires integrating over all latent variables $\latentvar$ or, we can use the chain rule of the probability;
\begin{align}
    \label{eq:chain_rule}
    \datadistribution(\sample) = \frac{\datadistribution(\sample,\latentvar)}{\datadistribution(\latentvar | \sample)}
\end{align}
which requires access to the ground truth encoder $\encodertrue(\latentvar|\sample)$. Both methods of maximizing the likelihood are intractable when using complex models.

\subsubsection{Evidence lower bound (ELBO).}

Instead of focusing on directly maximizing the likelihood $\datadistribution(\sample)$, we can work with the \emph{evidence}; the log-likelihood $\log\datadistribution(\sample)$. 
The ELBO represents the lower bound on the evidence. Let us build this lower bound in few steps. 
By definition ELBO is given by:   
\begin{align}
    \label{eq:elbo}
    \expectation_{\encoder(\latentvar|\sample)} \left[ \log \frac{\datadistribution(\sample,\latentvar)}{\encoder(\latentvar | \sample)} \right]
\end{align}
Looking at \cref{eq:chain_rule,eq:elbo}, we can derive the evidence in the form (see~\cite{luo2022understanding} for details):
\begin{align}
    \label{eq:evidence}
    \underbrace{\log \datadistribution(\sample)}_{\text{Evidence}} 
    = 
    \underbrace{\expectation_{\encoder(\latentvar|\sample)} \left[ \log \frac{\datadistribution(\sample,\latentvar)}{\encoder(\latentvar | \sample)} \right]}_{\text{ELBO}} 
    + 
    \underbrace{\kldivergence(\encoder(\latentvar|\sample) || \encodertrue(\latentvar|\sample))}_{\text{Distance}}
\end{align}
where $\kldivergence(\encoder(\latentvar|\sample) || \encodertrue(\latentvar|\sample))$ is a distance metric. 
This distance metric is the KL-divergence between the ground truth $\encodertrue(\latentvar|\sample)$ and our flexible approximate variational distribution $\encoder$ with parameters $\encoderparams$ that we seek to optimize. In other words, $\encoder$ seeks to approximate the true posterior $\encodertrue(\latentvar|\sample)$.

Since the second summand on the RHS of~\cref{eq:evidence} is a \emph{distance} metric, it is always positive. Therefore, the evidence is always higher than the ELBO term. In short, the evidence has a direct relationship wrt the ELBO term:
\begin{align}
    \log \datadistribution(\sample) \geq \expectation_{\encoder(\latentvar|\sample)} \left[ \log \frac{\datadistribution(\sample,\latentvar)}{\encoder(\latentvar | \sample)} \right]
\end{align}
which implies that the ELBO is the lower bound of the evidence. 

Note that the likelihood $\datadistribution(\sample)$ of our data---and therefore our evidence term $\log \datadistribution(\sample)$---is always a constant wrt $\encoderparams$,  as it is computed by marginalizing out all latents $\latentvar$ from the joint distribution $\datadistribution(\sample,\latentvar)$, thereby not depending on $\encoderparams$ whatsoever. Consequently, maximizing the ELBO would automatically result in minimizing the KL-divergence term on the RHS of~\cref{eq:evidence}, making $\encoder(\latentvar|\sample)$ better approximates the true posterior $\encodertrue(\latentvar|\sample)$.
Additionally, once trained, the ELBO can be used to estimate the likelihood of observed or generated data as well, since it is learned to approximate the model evidence $\log\datadistribution(\sample)$.

\subsubsection{Variational autoencoders (VAEs).}
An autoencoder is a type of neural network used to learn efficient codings of input data. 
It consists of two parts: 
%
{
\setlength{\columnsep}{4mm}%
\setlength{\intextsep}{4pt}%
\begin{wrapfigure}{r}{0pt}
    \includegraphics[width=0.2\textwidth]{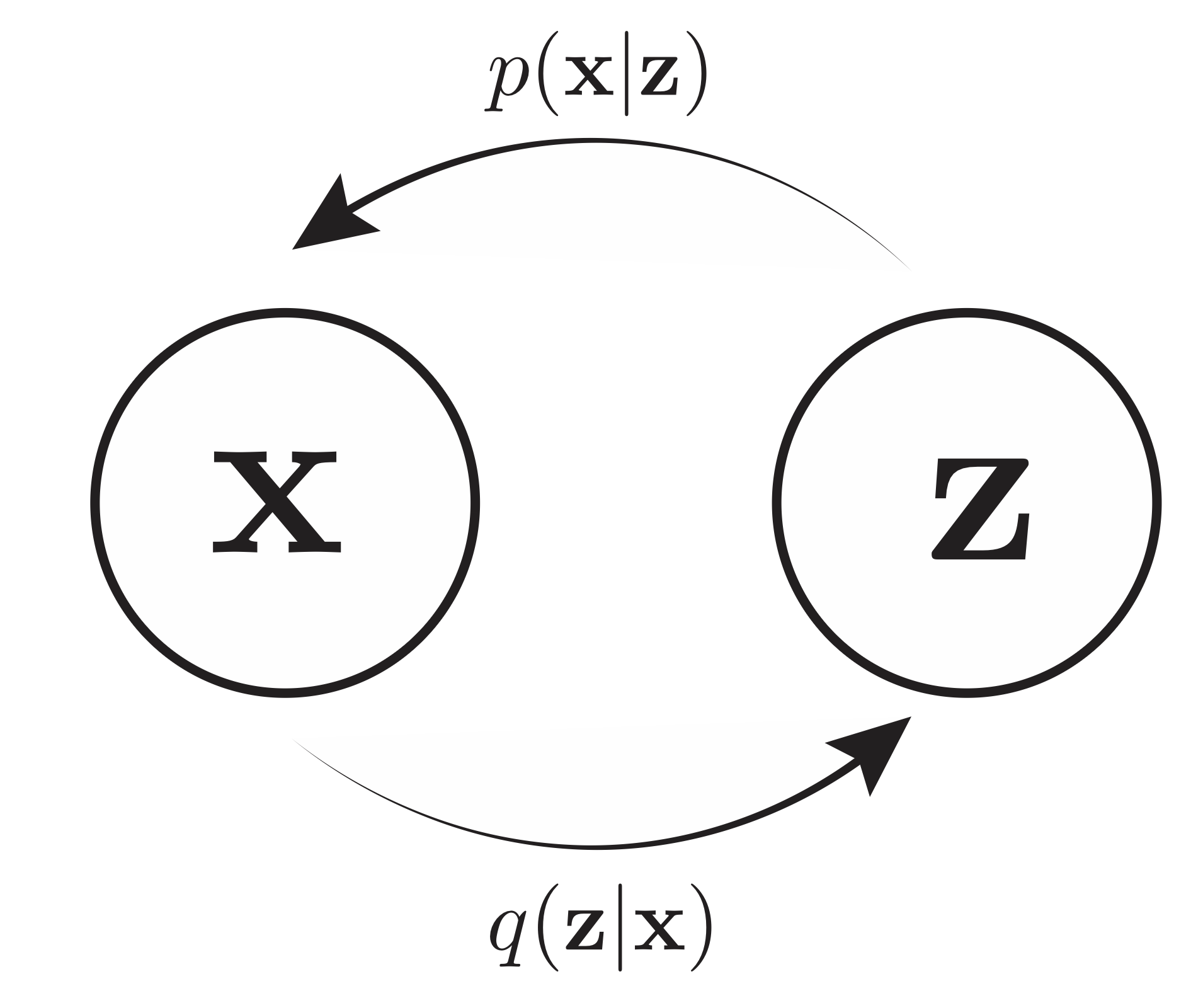}
\end{wrapfigure}
an encoder $\encoder(\latentvar | \sample)$ that maps the input data $\sample$ to a latent space $\latentvar$, and a decoder $\decoder(\sample | \latentvar)$ that maps the latent space back to the input data space. 
VAEs effectively maximizes the ELBO~\cref{eq:elbo}. The aproach is \emph{variational} as we optimizes our approximate posterior $\encoder$ by maximizing the ELBO. Mathematically speaking:
\begin{align}
    \label{eq:elbo_expand}
    \expectation_{\encoder(\latentvar|\sample)} \left[ \log \frac{\datadistribution(\sample,\latentvar)}{\encoder(\latentvar | \sample)} \right]
    &=
    \expectation_{\encoder(\latentvar|\sample)} \left[ \log \frac{\decoder(\sample|\latentvar)\datadistribution(\latentvar)}{\encoder(\latentvar | \sample)} \right]
    \;\; \text{using the Chain rule of Probability}
    \\
    &=  \expectation_{\encoder(\latentvar|\sample)}[\log \decoder(\sample|\latentvar)] + \expectation_{\encoder(\latentvar|\sample)}\left[\log\frac{\datadistribution(\latentvar)}{\encoder(\latentvar | \sample)}\right]
    \\
    &=  \underbrace{\expectation_{\encoder(\latentvar|\sample)}[\log \decoder(\sample|\latentvar)]}_{\text{reconstruction term}} 
    -\underbrace{\kldivergence(\encoder(\latentvar | \sample) || \datadistribution(\latentvar))}_{\text{prior matching term}}
\end{align}
}
%
In this case, we learn an intermediate bottleneck distribution $\encoder(\latentvar|\sample)$ that can be treated as an encoder ; it transforms inputs into a distribution over possible latents. Simultaneously, we learn a deterministic function $\decoder(\sample|\latentvar)$ to convert a given latent vector $\latentvar$ into an observation $\sample$, which can be interpreted as a decoder. The $\datadistribution(\latentvar)$ in the \emph{prior matching term} represents a known prior which is usually approximated to be a normal Gaussian distribution. 

\subsubsection{Hiearchical VAEs (HVAEs).}
HVAEs extend the concept of VAEs by incorporating multiple levels of latent variables, creating a hierarchy. This hierarchical structure allows HVAEs to model more complex data distributions and capture more intricate dependencies within the data. 
\begin{figure*}[!h]
    \centering
    \includegraphics[width=0.75\textwidth]{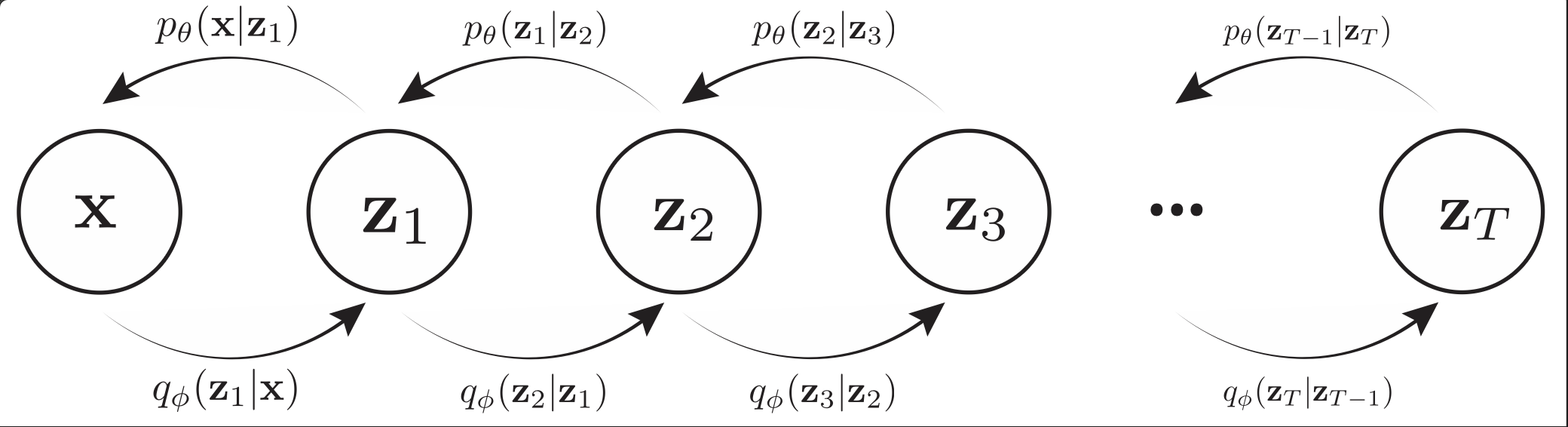}
    \caption{A Markovian Hiearchical VAE model with $\totaltime$ latent steps. The observed data $\sample$ goes through multiple levels of latent encodings $\latentvar_1,\ldots\latentvar_\totaltime$. In the generation process, each latent variable $\latentvar_t$ only depends on its previous level latent variable $\latentvar_{t-1}$.}
\end{figure*}
A standard VAE has a single layer of latent variables. The encoder maps the input data $\sample$ to a latent representation $\latentvar$, and the decoder reconstructs $\sample$ from $\latentvar$.
HVAEs introduce multiple layers of latent variables, organized hierarchically. The encoder produces a series of latent variables $\latentvar_1, \ldots, \latentvar_\totaltime$ at different levels. 
Each layer can capture different levels of abstraction, with higher layers capturing more abstract features and lower layers capturing more detailed features.
However, this increased expressiveness comes with added complexity in training and model architecture design. In this course, we focus only on Markovian HVAEs where the the generation process at each latent variable $\latentvar_t$ only depends on its previous level latent variable $\latentvar_{t-1}$.

\subsubsection{Variational diffusion models.}

The easiest way to think of a Variational Diffusion Model (VDM)~\citep{luo2022understanding,sohldickstein2015deep,ho2020ddpm,kingma2023variational} is simply as a Markovian Hierarchical Variational Autoencoder with three key restrictions:
\begin{itemize}
    \item The latent dimension is exactly equal to the data dimension
    \item The structure of the latent encoder at each timestep is not learned; it is pre-defined as a linear Gaussian model. In other words, it is a Gaussian distribution centered around the output of the previous timestep.
    \item The Gaussian parameters of the latent encoders vary over time in such a way that the distribution of the latent at final timestep $\totaltime$ is a standard Gaussian.
\end{itemize}
\begin{figure*}[!h]
    \centering
    \includegraphics[width=0.75\textwidth]{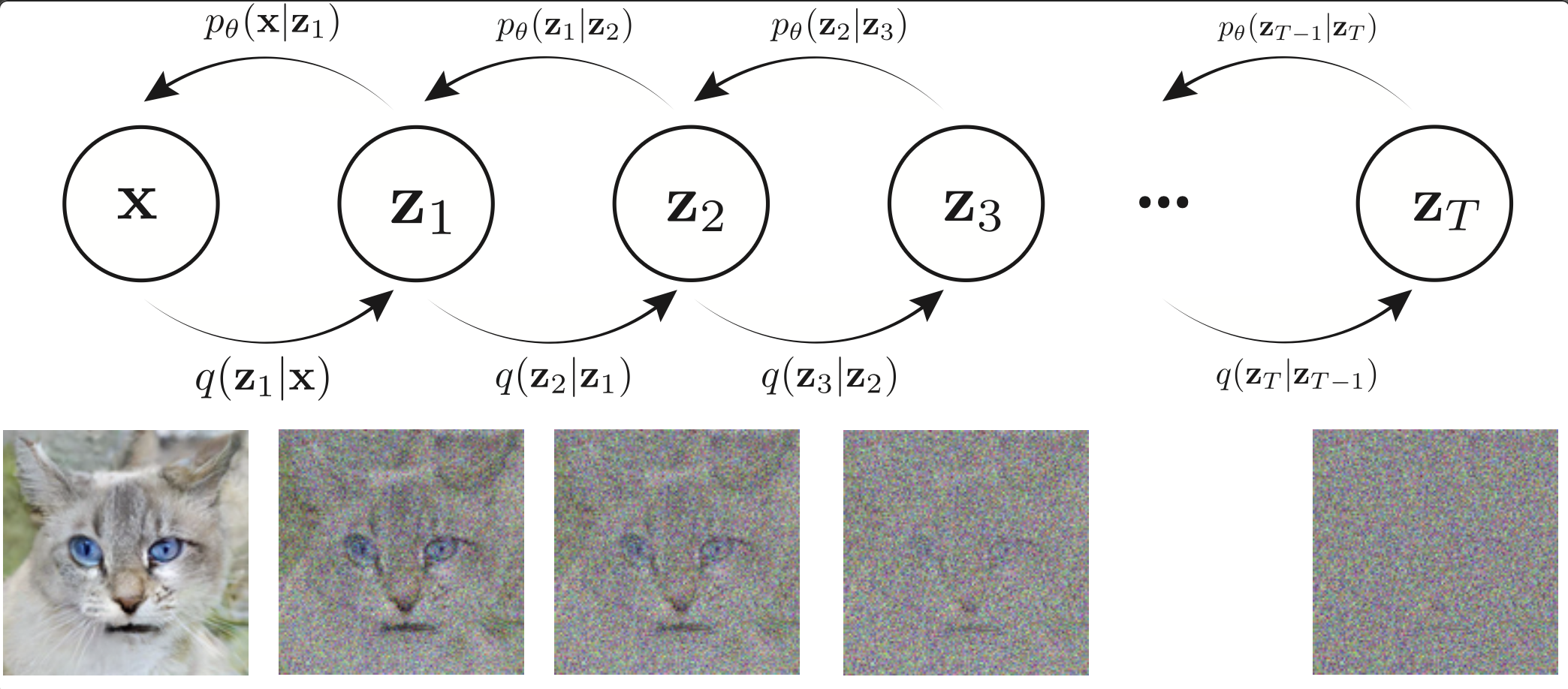}
    \caption{
    \label{fig:vdm}
    A variational diffusion model adds linear Gaussian noise at each time step, getting a fully Gaussian noise after $\totaltime$ latent steps.}
\end{figure*}
%
%
Note that our encoder distributions 
$\encoder(\latentvar_{t} | \latentvar_{t-1})$ in~\cref{fig:vdm} are no longer parameterized by $\encoderparams$, as they are completely modeled as Gaussians with defined mean and variance parameters at each timestep. 
Therefore, in a VDM, we are only interested in learning conditionals $\decoder(\latentvar_{t-1} | \latentvar_t)$, so that we can simulate new data. 
After optimizing the VDM, the sampling procedure is as simple as sampling Gaussian noise from $\decodertrue(\latentvar_\totaltime)$ and iteratively running the denoising transitions $\decoder(\latentvar_{t-1} | \latentvar_t)$ for $\totaltime$ steps to generate a novel $\sample$.


\paragraph{Connecting diffusion models with ELBO.}

Since in VDMs, the latent variables $\latentvar$ has the same dimensionality as the input, we represent them as $\sample_i$ with the input data sample as $\sample_0$. We can now resume from the log likelihood relation with ELBO as (see~\citep{luo2022understanding}):
\begin{align}
    \log \datadistribution(\sample) &\geq \expectation_{\encodertrue(\sample_{1:T}|\sample_0)} \left[ log \frac{\datadistribution(\sample_{0:T})}{\encodertrue (\sample_{1:T}| \sample_0)} \right] \\
    &= 
    \underbrace{\expectation_{\encodertrue(\sample_{1}|\sample_0)} [\log \decoder(\sample_0 | \sample_1)]}_{\text{reconstruction term}} 
    - 
    \underbrace{\kldivergence(\encodertrue(\sample_T|\sample_0) || \decodertrue(\sample_T))}_{\text{prior matching term}} 
    - \sum_{t=2}^T \underbrace{\expectation_{\encodertrue(\sample_{t}|\sample_0)} [\kldivergence(\encodertrue(\sample_{t-1}|\sample_t,\sample_0) \; || \; \decoder(\sample_{t-1}|\sample_t)) ].}_{\text{denoising matching term}}
\end{align}
Each term in this formulation has an elegant interpretation:
\begin{enumerate}
    \item The first term can be interpreted as a \emph{reconstruction term} and can be approximated and optimized using a Monte Carlo estimate
    \item The second term is a distance metric that tells how close the noisy version of the input $\sample_0$ to the standard Gaussian prior $\decodertrue(\sample_\totaltime)$. 
\end{enumerate}

While HVAEs and diffusion models are distinct in their methodologies and approaches to generative modeling, they share the common goal of learning complex data distributions. Their differences in hierarchical structures and generation processes provide unique strengths that, when combined, could lead to more advanced and capable generative models.

\subsection{Energy-based models (EBM)}
%
{
\setlength{\columnsep}{4mm}%
\setlength{\intextsep}{4pt}%
\begin{wrapfigure}{r}{0pt}
    \includegraphics[width=0.2\textwidth]{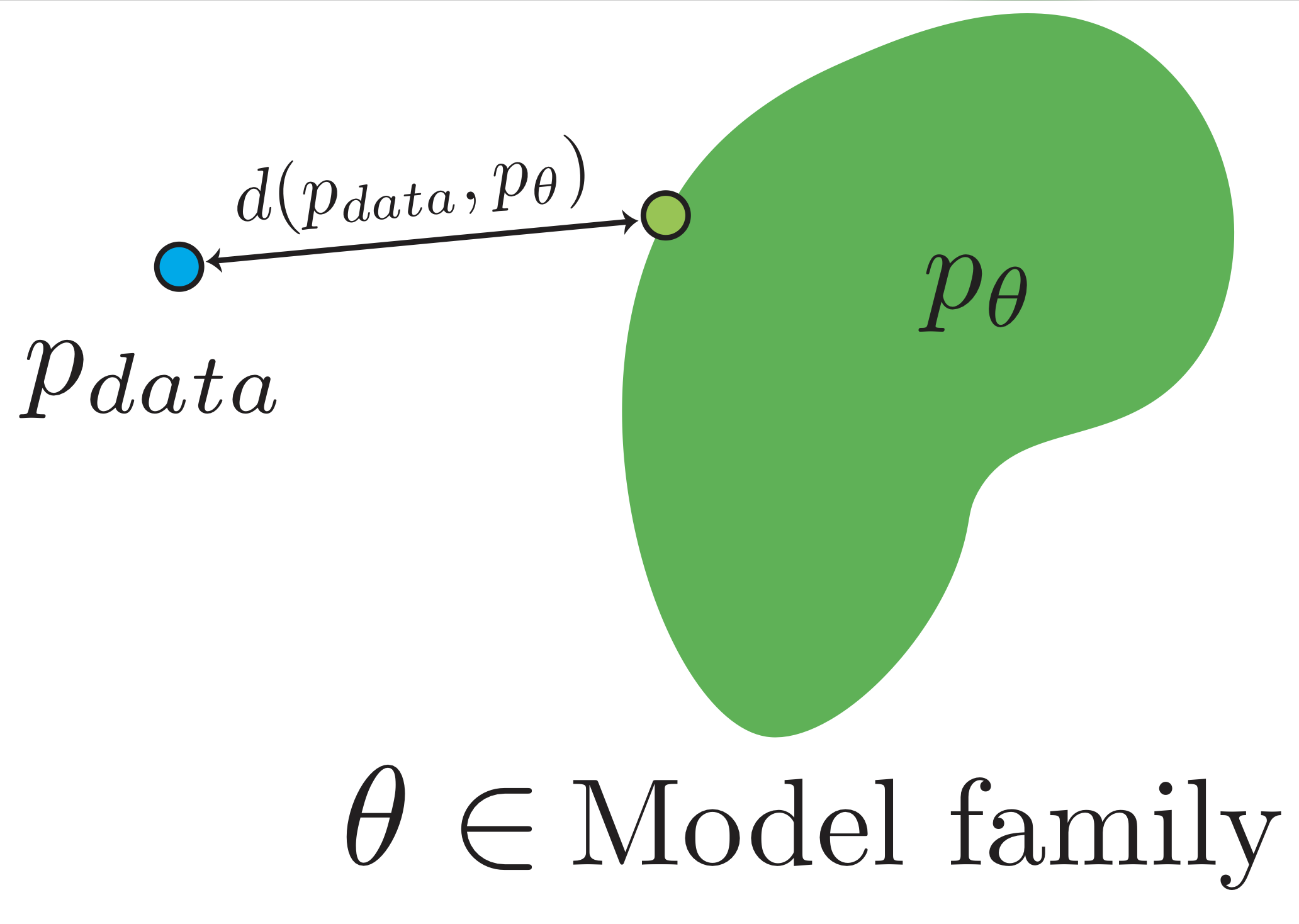}
\end{wrapfigure}
Another similar approach is energy-based
modeling, in which a distribution is learned as an arbitrarily flexible energy function that is then normalized.
EBMs are much less restrictive in functional form: instead of specifying a normalized probability, they only specify an unnormalized non-negatvive function of the form:
\begin{align}
    \label{eq:ebm_likelihood}
    \decoder(\sample) = \frac{\exp (\energy_\decoderparams(\sample))}{\normalization_\decoderparams}
    \text{\; where \;}
    \normalization_\decoderparams = \int \exp (\energy_\decoderparams(\sample)) d\sample
\end{align}
denotes the normalizing constant to ensure $\int \decoder(\sample) d\sample = 1$. 
$\energy_\decoderparams$ (the negative of the function $-\energy_\decoderparams$ is the energy) is a nonlinear regression function with parameters $\decoderparams$.  $\normalization_\decoderparams$ is constant wrt to $\sample$ and depends only on the parameters $\decoderparams$.
One way to learn such a distribution is by maximizing the likelihood. However, this requires computing the normalization constant $\normalization_\decoderparams$ which is intractable for complex energy functions.

With EBMs, we can side step such intractable computation by estimating the gradient of the log-likelihood using MCMC methods, which implicitly allows likelihood maximization with gradient ascent~\citep{younes2000convergence}.} 
Another important property of EBMs is that even though we cannot compute the likelihood of a sample, we can report the relative importance of any sample. This could be beneficial for many applications like object recognition, sequence labeling or image restoration. 
Given two samples $\sample$ and $\sample'$, the relative importance is the ratio:
\begin{align}
    \frac{\decoder(\sample)}{\decoder(\sample')} 
    = \frac{{\exp (\energy_\decoderparams(\sample))}/{\normalization_\decoderparams}}{{\exp (-\energy_\decoderparams(\sample'))}/{\normalization_\decoderparams}}
    = \frac{\exp (\energy_\decoderparams(\sample))}{\exp (-\energy_\decoderparams(\sample'))}
    = \exp(\energy_\decoderparams(\sample) - \energy_\decoderparams(\sample')).
\end{align}
Since the normalizing constant cancels out in the above ratio, we only need to deal with the $\exp \energy_\decoderparams(\cdot)$ terms which are tractable. In the next section, we see how MCMC methods can be used to train and sample from EBMs.

\paragraph{Contrastive Divergence algorithm.}
Intuitively, one can maximize the likelihood~\eqref{eq:ebm_likelihood} by increasing the numerator and decreasing the denominator. The contrastive divergence algorithm~\cite{hinton2002training} works on this contrastive idea. 
To maximize the log-likelihood, we optimize the parameters $\decoderparams$. This requires computing the gradient wrt  $\decoderparams$ which has the form:
\begin{align}
    \nabla_\decoderparams \log \decoder(\sample) 
    = \nabla_\decoderparams \log (\exp (\energy_\decoderparams(\sample)) /{\normalization_\decoderparams}) 
    &=
    \nabla_\decoderparams \energy_\decoderparams(\sample) - \nabla_\decoderparams \log \normalization_\decoderparams
    \\
    &=
    \nabla_\decoderparams \energy_\decoderparams(\sample) - \frac{\nabla_\decoderparams \normalization_\decoderparams}{\normalization_\decoderparams}
    \\
    &=
    \energy_\decoderparams(\sample) - \frac{1}{\normalization_\decoderparams} \nabla_\decoderparams \int \exp(\energy_\decoderparams(\sample)) d\sample
    \\
    &=
    \energy_\decoderparams(\sample) - \frac{1}{{{\normalization_\decoderparams} }} \int \exp(\energy_\decoderparams(\sample)) \nabla_\decoderparams \energy_\decoderparams(\sample) d\sample
    \\
    &=
    \energy_\decoderparams(\sample) - \int \frac{ \exp(\energy_\decoderparams(\sample))}{\normalization_\decoderparams} \nabla_\decoderparams \energy_\decoderparams(\sample) d\sample 
    \leftarrow  \text{the ratio is a pdf $\model(\sample)$} 
    \\
    \label{eq:expectation_score}
    &=
    \energy_\decoderparams(\sample) - \expectation_{\sample}[\nabla_\decoderparams \energy_\decoderparams(\sample) ]
    \\
    &= 
    \label{eq:one_sample_estimate}
    \energy_\decoderparams(\sample) - \nabla_\decoderparams \energy_\decoderparams(\sample_{\text{sample}}) \leftarrow  \text{one-sample estimate} 
\end{align}
The expectation term in~\cref{eq:expectation_score} can be approximated by Monte Carlo estimation. \Cref{eq:one_sample_estimate} is a one sample estimate of this expectation. One can sample $\sample_\text{sample} \sim \exp \energy(\sample_\text{sample})/\normalization_\decoderparams$ from the model $\model$ and take step in the direction given by the gradient $\nabla \exp(\energy_\decoderparams(\sample) - \energy_\decoderparams(\sample_\text{sample}))$ making training data more likely than a typical sample from the model. \emph{But the main question remains, how do we sample $\sample_\text{sample}$?} This is where MCMC methods comes into play.

\subsubsection{MCMC methods for sampling from EBMs}

EBMs are extremely flexible in the way $\energy_\decoderparams$ can be chosen. There is practically no restriction on the choice of $\energy_\decoderparams$. This means, you can plug-in whatever architecture you want to model the data. However, the problem is that sampling from $\model(\sample)$ is very hard. Generating new samples could be computationally very expensive from an EBM. 
The reason is that evaluating and optimizing likelihood $\model(\sample)$ is hard (\ie learning is hard). Even if you train your model $\model$, sampling would require finding the normalization constant $\normalization_\decoderparams$, which is intractable. This is because fundamentally the numerical cost to compute $\normalization_\decoderparams$ scales exponentially with the number of dimensions of $\sample$.

To optimize the log-likelihood form from~\cref{eq:one_sample_estimate}, we would like to pick samples $\sample_\text{sample}$ that represents the underlying data distribution. This can be achieved by using iterative algorithms. At each iteration, we can perform MCMC iterative sampling to obtain the sample $\sample_\text{sample}$ which can be used to evaluate~\cref{eq:one_sample_estimate}. We discuss two important MCMC algorithms below.

\paragraph{Metropolis-Hastings MCMC}
We can use Metropolis-Hastings MCMC sampling algorithm which is an iterative method:
\begin{itemize}
    \item Initialize a random sample $\sample_0$ at $t=0$
    \item Repeat the process for $t = 0,1,2,\cdots,\totaltime-1$
    \subitem $\sample' = \sample^t + \text{noise}$
    \subsubitem If $\energy(\sample') > \energy(\sample^t)$ then $\sample^{t+1} = \sample' $
    \subsubitem else let $\energy(\sample^{t+1}) = \sample'$ with probability $\exp (\energy_\decoderparams(\sample^t) - \energy_\decoderparams(\sample'))$
\end{itemize}
Here the noise term could be any perturbation that can be added to the data sample. Note that, unlike standard MH algorithm---where the samples rejected if they fall below the accepted probability---we accept samples which have lower probability than the acceptance ratio. This is because we want to keep sampling the space. We simply ensure tehe sample is accepted with the lower probaility. In theory, MCMC sampling works but it can take quite a long time to converge.

\paragraph{Unadjusted Langevin MCMC}
%
Slightly better version of MCMC sampling is using Langevin dynamics. Sampling using the unadjusted Langevin MCMC algorithm (ULA) works as follows:
\begin{itemize}
    \item Initialize a random sample $\sample^0 \sim \targetdistribution(\sample)$ at $t=0$
    \item Repeat for $t = 0,1,2,\ldots, \totaltime-1$
    \subitem $\latentvar^t \sim \normaldistribution(0,I)$
    \subitem $\sample^{t+1} = \sample^t + \epsilon \nabla_x \log \decoder(\sample)|_{\sample=\sample^t} + \sqrt{2\epsilon} \latentvar^t$
\end{itemize}
Here $\epsilon$ is the step size. Note that ULA has no rejection step. The samples are perturbed with a noise $\latentvar_t$ and the new sample is accepted. However, in theory this algorithm can only converge if the step size $\epsilon$ is small. But this can slow down the convergence dramatically. At least, the perturbation is informed thanks to the $\nabla_x \decoder(\sample)|_{\sample=\sample_t}$ term. In MH algorithm, the noise added could be any random perturbation. ULA has many noticeable properties:
\begin{itemize}
    \item $\sample^\totaltime$ converges to a sample from $\decoder(\sample)$ as $\totaltime \rightarrow \infty$ and the step size $\epsilon \rightarrow 0$. 
    \item ULA has better convergence compared to MH MCMC because now the exploration is informed thanks to the $\nabla_\sample \log \decoder(\sample)$ term. 
    \item $\nabla_\sample \log \decoder(\sample) = \nabla_x\energy_\decoderparams(\sample)$ for continuous energy models, which implies that the normalization term is completely gone. 
\end{itemize}
There are still issues with this approach. Sampling converges slowly in higher dimensional spaces and is thus very expensive, yet we need sampling in every iteration in contrastive divergence. 
For example, every time we need to sample to solve~\cref{eq:one_sample_estimate}, we need to run (say 1000) large number of iterations of an MCMC algorithm to generate a valid sample that follows the underlying distribution $\decoder$.Yet, we need sampling \emph{at each iteration} in contrastive divergence.
\cite{gao2021learning} propose to train energy-based models by diffusion recovery likelihood where long-run MCMC samples from the conditional distributions do not diverge and still represent realistic images. This allows them to accurately estimate the normalized density of data even for high-dimensional datasets.

\paragraph{Training without sampling.}

To avoid such expensive sampling procedures one can use training methods that does not require sampling. Score matching~\citep{hyvarinen2005estimation,song2020generative,song2019sliced} and noise contrast estimation~\citep{gutmann2010noise} algorithms are such methods that does not require sampling for training. We will not delve much into these methods. However, we will briefly looking score-based models in the next section as they require sampling to generate new samples where MCMC methods shine again.

\subsection{Sampling score-based generative models}

So far we have been talking about energy-based models which suffer from expensive sampling during the training stage. Score-based models can help avoid this sampling step altogether from the training stage. 
Score-based generative models are highly related to EBMs; instead of learning to model the energy function itself, they
learn the score of the energy-based model as a neural network. 

Starting from~\cref{eq:ebm_likelihood}:
\begin{align}
    \model(\sample) &= \frac{\exp \energy_\decoderparams(\sample)}{\normalization_\decoderparams} 
    \\
    \log \model(\sample) &= \energy_\decoderparams(\sample) - \log \normalization_\decoderparams
    \\
    \gradientx \log \model(\sample) &= \gradientx 
 \energy_\decoderparams(\sample) - \underbrace{\gradientx  \log \normalization_\decoderparams}_{=0}
 \\
    \gradientx \log \model(\sample) &= \gradientx 
 \energy_\decoderparams(\sample) = \score(\sample)
\end{align}
Taking the gradient of the log-likelihood wrt to $\sample$ renders the summand on the RHS to zero since the normalization constant (aka the partion function) $\normalization_\decoderparams$ only depends on $\decoderparams$. The $\score(\sample)$ is the \emph{score function}.

The \emph{score} $\score(\sample)$ provides an alternative view of the original function where you are looking at things from the perspective of the gradient instead of the perspective of the likelihood itself. The key observation here is that the score is independent of the normalization constant (aka the partion function) $\normalization_\decoderparams$.

What does the score function represent? For every $\sample$, taking the gradient of its log likelihood with respect
to $\sample$ essentially describes what direction in data space to move in order to further increase its likelihood.
Intuitively, the score function defines a vector field over the entire data space, pointing towards the modes.


Some efficient MCMC methods, such
as Langevin MCMC or Hamiltonian MC~\cite{neal2012hmc}, make use of the fact that the gradient of the log-probability wrt $\sample$ (a.k.a \emph{score}) is equal to the (negative) gradient of the energy, therefore easy to calculate.


By learning the score function of the true data distribution, we can generate samples by starting at any arbitrary point in the same space and iteratively following the score until a mode is reached. This sampling procedure is known as Langevin dynamics, and is mathematically described as:
\begin{align}
    \sample_{i+1} = \sample_i + c \nabla \log\datadistribution(\sample_i) + \sqrt{2c}\epsilon
\end{align}
Collectively, learning to represent a distribution as a score function and using it to generate samples through MCMC techniques, such as Langevin dynamics, is known as Score-based Generative Modeling~\citep{song2021score,song2020generative,song2020improved}.

\section{Conclusion}
MCMC methods offer a unified framework for sampling from complex probability distributions, addressing a common challenge across the domains of rendering, generative modeling, and optimization. Significant research efforts have been dedicated to identifying the conceptual bridge between these interconnected fields. 

However, there are a number of challenges that need to be addressed. For example, the chain constructed by the Ordinary Metropolis Hastings algorithm is not only invariant, but even reversible with respect to the target distribution. This invariance property, which is necessary for usage in Monte Carlo sampling, highly restricts our choice of processes, which we can use for state space exploration. That is, even when we have a process at hand, which is able to explore our given state space in
a favorable manner, we cannot use it for Monte Carlo sampling, unless it is invariant. 
Reversibility is an even stronger condition. While it yields certain useful spectral properties of the process, it also slows down mixing and convergence to equilibrium. Reversible processes show backtracking behavior where the processes frequently revisit previously visited states before reaching unexplored areas, thereby, slowing
down the convergence. 
Another challenge with designing MCMC algorithms is that the
highly correlated Markov chains can cause excessive local exploration which in rendering are visible as overly bright pixels. To avoid
this issue, MCMC algorithms should discover other contributing
areas by globally discovering the paths away from the current path.
All existing algorithms~\citep{veach1997metropolis,luan2020langevin,li2015anistropic,Pantaleoni_2017,Bitterli2017reversible,bitterli19selectively} are based Metropolis-Hastings which inherits its
slower convergence due to the reversibility property. 
Recently, \cite{holl2024jump} introduced a continuous time Markov framework that is based on the restore algorithm~\citep{wang2021regeneration}. Their framework  adjusts an arbitrary Markov
chain for Monte Carlo integration to the graphics community. Especially, they introduced a rejection-free and target density sensitive way to deal with global discovery. They generalized the idea
presented in~\cite{wang2021regeneration} and extended it for light transport rendering problems. 
Given the surge in generative models, this work provides a solid foundational framework that can be leveraged to establish direct connections between diffusion models in generative modeling, physically based rendering and SGD-based optimization algorithms. 

We believe that this course will equip participants with the knowledge to bridge the gaps between physically based rendering and vision-based generative modeling more effectively. We are excited about how understanding MCMC will enable participants to recognize the common probabilistic foundation underlying diverse applications. This insight will enhance their ability to apply learned concepts across different domains.

\section{Speakers}
\paragraph{Gurprit Singh} is a senior researcher at the Max Planck Institute (MPI) for Informatics, where he is currently leading the sampling \& rendering group. His research interests are in Markov Chain Monte Carlo (MCMC), diffusion models, noise correlations and Monte Carlo sampling for physically based (forward/inverse) rendering. He has published his research in the top-tier conferences like SIGGRAPH (North America/Asia), NeurIPS, ECCV, Eurographics. 
His current focus is on bridging the gap between physically based rendering and generative AI using the powerful machinery of MCMC methods.
\newline
Email: gsingh@mpi-inf.mpg.de \qquad Webpage: \href{https://sampling.mpi-inf.mpg.de/}{https://sampling.mpi-inf.mpg.de/}

\paragraph{Wenzel Jakob} is an associate professor leading the Realistic Graphics Lab at EPFL's School of Computer and Communication Sciences. Wenzel has received the ACM SIGGRAPH Significant Researcher award, the Eurographics Young Researcher Award, and an ERC Starting Grant. His group develops the Mitsuba renderer~\cite{mitsuba3}, a research-oriented rendering system, and he has co-authored the third and fourth editions of Physically Based Rendering: From Theory To Implementation~\cite{pharr2023physically}.
\newline
Email: wenzel.jakob@epfl.ch \qquad
Webpage: \href{https://rgl.epfl.ch/people/wjakob}{https://rgl.epfl.ch/people/wjakob}

\appendix
\section{Code snippets}
\label{sec:coding}

\begin{lstlisting}[language=Python, caption=Python example]
# Brownian motion
import numpy as np
import matplotlib.pyplot as plt

# Function to simulate Brownian motion
def brownian_motion(num_steps, step_size):
    # Initialize position at the origin
    position = np.zeros(2)
    positions = [position.copy()]

    # Iterate through each step
    for _ in range(num_steps):
        # Generate random displacement for each dimension
        displacement = np.random.normal(0, step_size, size=2)
        # Update position
        position += displacement
        # Store the updated position
        positions.append(position.copy())

    return np.array(positions)

# Number of steps and step size
num_steps = 1000
step_size = 0.1

# Simulate Brownian motion
positions = brownian_motion(num_steps, step_size)

# Plot the Brownian motion trajectory
plt.plot(positions[:, 0], positions[:, 1], lw=1)
plt.title('2D Brownian Motion')
plt.xlabel('X')
plt.ylabel('Y')
plt.show()

\end{lstlisting}

\begin{lstlisting}[language=Python, caption=Python example]

# Unadjusted Langevin Monte Carlo sampling
import numpy as np
import matplotlib.pyplot as plt

def target_log_prob(x):
    Example target distribution: standard normal distribution

def grad_target_log_prob(x):
    # Gradient of the log of the target distribution: standard normal distribution
    return -x

def langevin_monte_carlo(num_samples, dim, step_size, burn_in):
    samples = np.zeros((num_samples, dim))
    current_sample = np.random.randn(dim)  # Initialize with a random sample
    
    for i in range(num_samples + burn_in):
        grad_log_prob = grad_target_log_prob(current_sample)
        noise = np.random.randn(dim)
        next_sample = current_sample + 0.5 * step_size * grad_log_prob + np.sqrt(step_size) * noise
        
        # Accept the next sample
        current_sample = next_sample
        
        if i >= burn_in:
            samples[i - burn_in] = current_sample
            
    return samples

# Parameters
num_samples = 10000  # Number of samples to generate
dim = 2            # Dimension of the target distribution
step_size = 0.1    # Step size for the Langevin dynamics
burn_in = 100      # Number of burn-in steps

# Generate samples
samples = langevin_monte_carlo(num_samples, dim, step_size, burn_in)


# Create subplots
fig, axs = plt.subplots(1, 1, figsize=(5, 5))

# Plot Brownian motion trajectory
# axs[0].plot(positions[:, 0], positions[:, 1], lw=1)
axs.scatter(samples[:, 0], samples[:, 1], s=0.5)
axs.set_title('LMC')
axs.set_xlabel('X')
axs.set_ylabel('Y')
axs.set_xlim(-3, 3)
axs.set_ylim(-3, 3)
# axs.grid(True)
axs.set_aspect(1)

plt.tight_layout()
plt.show()
\end{lstlisting}

\begin{lstlisting}[language=Python, caption=Python example]

# Unadjusted Hamiltonian Monte Carlo
import numpy as np
import matplotlib.pyplot as plt

# Define the target distribution (posterior) - For example, a Gaussian distribution
def log_prob(theta):
    """Returns the log of the target distribution (up to a constant)"""
    return -0.5 * np.sum(theta ** 2)

# Gradient of the log-probability
def grad_log_prob(theta):
    """Returns the gradient of the log-probability"""
    return -theta

# Hamiltonian Monte Carlo Sampling
def hmc(log_prob, grad_log_prob, initial_theta, n_samples, step_size, n_leapfrog):
    """
    log_prob: function to compute log-probability of the target distribution
    grad_log_prob: function to compute the gradient of the log-probability
    initial_theta: initial value of parameters (starting point)
    n_samples: number of samples to generate
    step_size: step size for the leapfrog integrator
    n_leapfrog: number of leapfrog steps in the simulation
    """
    samples = []
    current_theta = initial_theta
    current_log_prob = log_prob(current_theta)
    
    for _ in range(n_samples):
        % # Sample a random momentum (p) from a normal distribution
        current_p = np.random.normal(0, 1, size=current_theta.shape)
        initial_p = current_p
        
        % # Hamiltonian dynamics step (Leapfrog integrator)
        theta = np.copy(current_theta)
        p = np.copy(current_p)
        
        % # Half-step update of momentum
        p -= 0.5 * step_size * grad_log_prob(theta)
        
        % # Full-step updates of position (theta) and momentum (p)
        for _ in range(n_leapfrog):
            # Update theta
            theta += step_size * p
            
            % # Update momentum (except the last iteration)
            if _ < n_leapfrog - 1:
                p -= step_size * grad_log_prob(theta)
        
        % # Final half-step update of momentum
        p -= 0.5 * step_size * grad_log_prob(theta)
        
        % # Negate the momentum to make the proposal symmetric
        p = -p
        
        % # Compute Hamiltonian at the start and end of the trajectory
        current_H = -current_log_prob + 0.5 * np.sum(initial_p ** 2)
        proposed_H = -log_prob(theta) + 0.5 * np.sum(p ** 2)
        
        % # Metropolis acceptance criterion
        if np.random.uniform(0, 1) < np.exp(current_H - proposed_H):
            current_theta = theta
            current_log_prob = log_prob(current_theta)
        
        samples.append(current_theta)
    
    return np.array(samples)

# Generate samples using HMC
samples = hmc(log_prob, grad_log_prob, initial_theta, n_samples, step_size, n_leapfrog)
\end{lstlisting}



\bibliography{main}

\begin{thebibliography}{35}
\providecommand{\natexlab}[1]{#1}
\providecommand{\url}[1]{\texttt{#1}}
\expandafter\ifx\csname urlstyle\endcsname\relax
  \providecommand{\doi}[1]{doi: #1}\else
  \providecommand{\doi}{doi: \begingroup \urlstyle{rm}\Url}\fi

\bibitem[Betancourt(2018)]{betancourt2018conceptual}
M.~Betancourt.
\newblock A conceptual introduction to hamiltonian monte carlo, 2018.
\newblock URL \url{https://arxiv.org/abs/1701.02434}.

\bibitem[Bitterli and Jarosz(2019)]{bitterli19selectively}
B.~Bitterli and W.~Jarosz.
\newblock Selectively {{Metropolised}} {{Monte}} {{Carlo}} light transport
  simulation.
\newblock \emph{ACM Transactions on Graphics (Proceedings of SIGGRAPH Asia)},
  38\penalty0 (6), Nov. 2019.
\newblock \doi{10.1145/3355089.3356578}.

\bibitem[Bitterli et~al.(2017)Bitterli, Jakob, Nov{\'a}k, and
  Jarosz]{Bitterli2017reversible}
B.~Bitterli, W.~Jakob, J.~Nov{\'a}k, and W.~Jarosz.
\newblock Reversible jump metropolis light transport using inverse mappings.
\newblock \emph{ACM Transactions on Graphics}, 37\penalty0 (1), Oct. 2017.
\newblock \doi{10.1145/3132704}.

\bibitem[Chen et~al.(2016)Chen, Carlson, Gan, Li, and Carin]{chen2016bridging}
C.~Chen, D.~Carlson, Z.~Gan, C.~Li, and L.~Carin.
\newblock {Bridging the Gap between Stochastic Gradient MCMC and Stochastic
  Optimization}.
\newblock In A.~Gretton and C.~C. Robert, editors, \emph{Proceedings of the
  19th International Conference on Artificial Intelligence and Statistics},
  volume~51 of \emph{Proceedings of Machine Learning Research}, pages
  1051--1060, Cadiz, Spain, 09--11 May 2016. PMLR.
\newblock URL \url{https://proceedings.mlr.press/v51/chen16c.html}.

\bibitem[Gao et~al.(2021)Gao, Song, Poole, Wu, and Kingma]{gao2021learning}
R.~Gao, Y.~Song, B.~Poole, Y.~N. Wu, and D.~P. Kingma.
\newblock Learning energy-based models by diffusion recovery likelihood, 2021.
\newblock URL \url{https://arxiv.org/abs/2012.08125}.

\bibitem[Gutmann and Hyvärinen(2010)]{gutmann2010noise}
M.~Gutmann and A.~Hyvärinen.
\newblock Noise-contrastive estimation: A new estimation principle for
  unnormalized statistical models.
\newblock In Y.~W. Teh and M.~Titterington, editors, \emph{Proceedings of the
  Thirteenth International Conference on Artificial Intelligence and
  Statistics}, volume~9 of \emph{Proceedings of Machine Learning Research},
  pages 297--304, Chia Laguna Resort, Sardinia, Italy, 13--15 May 2010. PMLR.
\newblock URL \url{https://proceedings.mlr.press/v9/gutmann10a.html}.

\bibitem[Hairer et~al.(2013)Hairer, Lubich, and Wanner]{hairer2013geometric}
E.~Hairer, C.~Lubich, and G.~Wanner.
\newblock \emph{Geometric Numerical Integration: Structure-Preserving
  Algorithms for Ordinary Differential Equations}.
\newblock Springer Series in Computational Mathematics. Springer Berlin
  Heidelberg, 2013.
\newblock ISBN 9783662050187.
\newblock URL \url{https://books.google.cz/books?id=cPTxCAAAQBAJ}.

\bibitem[Hinton(2002)]{hinton2002training}
G.~E. Hinton.
\newblock Training products of experts by minimizing contrastive divergence.
\newblock \emph{Neural Comput.}, 14\penalty0 (8):\penalty0 1771–1800, aug
  2002.
\newblock ISSN 0899-7667.
\newblock \doi{10.1162/089976602760128018}.
\newblock URL \url{https://doi.org/10.1162/089976602760128018}.

\bibitem[Ho et~al.(2020)Ho, Jain, and Abbeel]{ho2020ddpm}
J.~Ho, A.~Jain, and P.~Abbeel.
\newblock Denoising diffusion probabilistic models.
\newblock In \emph{Proceedings of the 34th International Conference on Neural
  Information Processing Systems}, NIPS '20, Red Hook, NY, USA, 2020. Curran
  Associates Inc.
\newblock ISBN 9781713829546.

\bibitem[Holl et~al.(2024)Holl, Seidel, and Singh]{holl2024jump}
S.~Holl, H.-P. Seidel, and G.~Singh.
\newblock Jump restore light transport, 2024.
\newblock URL \url{https://arxiv.org/abs/2409.07148}.

\bibitem[Hyv{{\"a}}rinen(2005)]{hyvarinen2005estimation}
A.~Hyv{{\"a}}rinen.
\newblock Estimation of non-normalized statistical models by score matching.
\newblock \emph{Journal of Machine Learning Research}, 6\penalty0
  (24):\penalty0 695--709, 2005.
\newblock URL \url{http://jmlr.org/papers/v6/hyvarinen05a.html}.

\bibitem[Ivo et~al.(2019)Ivo, V{\'{e}}voda, Grittmann, Sk{\v{r}}ivan,
  Slusallek, and K{\v{r}}iv{\'{a}}nek]{Kondapaneni2019}
K.~Ivo, P.~V{\'{e}}voda, P.~Grittmann, T.~Sk{\v{r}}ivan, P.~Slusallek, and
  J.~K{\v{r}}iv{\'{a}}nek.
\newblock Optimal multiple importance sampling.
\newblock \emph{ACM Transactions on Graphics (Proceedings of SIGGRAPH 2019)},
  38\penalty0 (4):\penalty0 37:1--37:14, July 2019.
\newblock \doi{10.1145/3306346.3323009}.

\bibitem[Jakob et~al.(2022)Jakob, Speierer, Roussel, Nimier-David, Vicini,
  Zeltner, Nicolet, Crespo, Leroy, and Zhang]{mitsuba3}
W.~Jakob, S.~Speierer, N.~Roussel, M.~Nimier-David, D.~Vicini, T.~Zeltner,
  B.~Nicolet, M.~Crespo, V.~Leroy, and Z.~Zhang.
\newblock {Mitsuba 3 renderer}, 2022.
\newblock URL \url{https://mitsuba-renderer.org}.

\bibitem[Kelemen et~al.(2002)Kelemen, Szirmay-Kalos, Antal, and
  Csonka]{Kelemen:2002:ASA}
C.~Kelemen, L.~Szirmay-Kalos, G.~Antal, and F.~Csonka.
\newblock A simple and robust mutation strategy for the {Metropolis} light
  transport algorithm.
\newblock \emph{Computer Graphics Forum}, 21\penalty0 (3):\penalty0 531--540,
  2002.

\bibitem[Kingma et~al.(2023)Kingma, Salimans, Poole, and
  Ho]{kingma2023variational}
D.~P. Kingma, T.~Salimans, B.~Poole, and J.~Ho.
\newblock Variational diffusion models, 2023.
\newblock URL \url{https://arxiv.org/abs/2107.00630}.

\bibitem[Kollig and Keller(2002)]{kollig2002efficient}
T.~Kollig and A.~Keller.
\newblock \emph{Efficient bidirectional path tracing by randomized quasi-{Monte
  Carlo} integration}.
\newblock Springer, 2002.

\bibitem[Leimkuhler and Reich(2005)]{leimkuhler2005hamiltonian}
B.~Leimkuhler and S.~Reich.
\newblock \emph{Simulating Hamiltonian Dynamics}.
\newblock Cambridge Monographs on Applied and Computational Mathematics.
  Cambridge University Press, 2005.

\bibitem[Li et~al.(2015)Li, Lehtinen, Ramamoorthi, Jakob, and
  Durand]{li2015anistropic}
T.-M. Li, J.~Lehtinen, R.~Ramamoorthi, W.~Jakob, and F.~Durand.
\newblock {Anisotropic Gaussian Mutations for Metropolis Light Transport
  through Hessian-Hamiltonian Dynamics}.
\newblock \emph{ACM Trans. Graph.}, 34\penalty0 (6), nov 2015.
\newblock ISSN 0730-0301.
\newblock \doi{10.1145/2816795.2818084}.
\newblock URL \url{https://doi.org/10.1145/2816795.2818084}.

\bibitem[Luan et~al.(2020)Luan, Zhao, Bala, and Gkioulekas]{luan2020langevin}
F.~Luan, S.~Zhao, K.~Bala, and I.~Gkioulekas.
\newblock {Langevin Monte Carlo Rendering with Gradient-based Adaptation}.
\newblock \emph{ACM Trans. Graph.}, 39\penalty0 (4), 2020.
\newblock URL \url{https://dl.acm.org/doi/abs/10.1145/3386569.3392382}.

\bibitem[Luo(2022)]{luo2022understanding}
C.~Luo.
\newblock {Understanding Diffusion Models: A Unified Perspective}, 2022.
\newblock URL \url{https://arxiv.org/abs/2208.11970}.

\bibitem[Pantaleoni(2017)]{Pantaleoni_2017}
J.~Pantaleoni.
\newblock Charted metropolis light transport.
\newblock \emph{ACM Transactions on Graphics}, 36\penalty0 (4):\penalty0
  1–14, July 2017.
\newblock ISSN 1557-7368.
\newblock \doi{10.1145/3072959.3073677}.
\newblock URL \url{http://dx.doi.org/10.1145/3072959.3073677}.

\bibitem[Pharr et~al.(2023)Pharr, Jakob, and Humphreys]{pharr2023physically}
M.~Pharr, W.~Jakob, and G.~Humphreys.
\newblock \emph{{Physically based rendering: From theory to implementation}}.
\newblock Morgan Kaufmann, 2023.
\newblock URL \url{https://www.pbr-book.org/}.

\bibitem[Radford(2011)]{neal2012hmc}
N.~Radford.
\newblock Mcmc using hamiltonian dynamics.
\newblock May 2011.
\newblock \doi{10.1201/b10905}.
\newblock URL \url{https://arxiv.org/pdf/1206.1901}.

\bibitem[Roberts and Stramer(2002)]{roberts2002langevin}
G.~O. Roberts and O.~Stramer.
\newblock {Langevin Diffusions and Metropolis-Hastings Algorithms}.
\newblock \emph{Method. Comput. Appl. Prob.}, 4\penalty0 (4):\penalty0
  337–357, dec 2002.
\newblock ISSN 1387-5841.
\newblock \doi{10.1023/A:1023562417138}.
\newblock URL \url{https://doi.org/10.1023/A:1023562417138}.

\bibitem[Sohl-Dickstein et~al.(2015)Sohl-Dickstein, Weiss, Maheswaranathan, and
  Ganguli]{sohldickstein2015deep}
J.~Sohl-Dickstein, E.~A. Weiss, N.~Maheswaranathan, and S.~Ganguli.
\newblock {Deep Unsupervised Learning using Nonequilibrium Thermodynamics},
  2015.
\newblock URL \url{https://arxiv.org/abs/1503.03585}.

\bibitem[Song and Ermon(2020{\natexlab{a}})]{song2020generative}
Y.~Song and S.~Ermon.
\newblock Generative modeling by estimating gradients of the data distribution,
  2020{\natexlab{a}}.
\newblock URL \url{https://arxiv.org/abs/1907.05600}.

\bibitem[Song and Ermon(2020{\natexlab{b}})]{song2020improved}
Y.~Song and S.~Ermon.
\newblock Improved techniques for training score-based generative models.
\newblock In \emph{Proceedings of the 34th International Conference on Neural
  Information Processing Systems}, NIPS '20, Red Hook, NY, USA,
  2020{\natexlab{b}}. Curran Associates Inc.
\newblock ISBN 9781713829546.

\bibitem[Song et~al.(2019)Song, Garg, Shi, and Ermon]{song2019sliced}
Y.~Song, S.~Garg, J.~Shi, and S.~Ermon.
\newblock Sliced score matching: A scalable approach to density and score
  estimation, 2019.
\newblock URL \url{https://arxiv.org/abs/1905.07088}.

\bibitem[Song et~al.(2021)Song, Sohl-Dickstein, Kingma, Kumar, Ermon, and
  Poole]{song2021score}
Y.~Song, J.~Sohl-Dickstein, D.~P. Kingma, A.~Kumar, S.~Ermon, and B.~Poole.
\newblock {Score-Based Generative Modeling through Stochastic Differential
  Equations}, 2021.
\newblock URL \url{https://arxiv.org/abs/2011.13456}.

\bibitem[Veach(1998)]{veach1998thesis}
E.~Veach.
\newblock \emph{{Robust {Monte Carlo} Methods for Light Transport Simulation}}.
\newblock PhD thesis, Stanford University, Stanford, CA, USA, 1998.
\newblock AAI9837162.

\bibitem[Veach and Guibas(1995)]{Veach:1995:OCS}
E.~Veach and L.~J. Guibas.
\newblock Optimally combining sampling techniques for {M}onte {C}arlo
  rendering.
\newblock In \emph{Proceedings of the 22nd annual conference on Computer
  graphics and interactive techniques}, SIGGRAPH '95, pages 419--428. ACM,
  1995.

\bibitem[Veach and Guibas(1997)]{veach1997metropolis}
E.~Veach and L.~J. Guibas.
\newblock {Metropolis light transport}.
\newblock In \emph{Proceedings of the 24th Annual Conference on Computer
  Graphics and Interactive Techniques}, SIGGRAPH '97, page 65–76, USA, 1997.
  ACM Press/Addison-Wesley Publishing Co.
\newblock ISBN 0897918967.
\newblock \doi{10.1145/258734.258775}.
\newblock URL \url{https://doi.org/10.1145/258734.258775}.

\bibitem[Wang et~al.(2021)Wang, Pollock, Roberts, and
  Steinsaltz]{wang2021regeneration}
A.~Q. Wang, M.~Pollock, G.~O. Roberts, and D.~Steinsaltz.
\newblock Regeneration-enriched markov processes with application to monte
  carlo.
\newblock \emph{The Annals of Applied Probability}, 31\penalty0 (2), Apr. 2021.
\newblock ISSN 1050-5164.
\newblock \doi{10.1214/20-aap1602}.
\newblock URL \url{http://dx.doi.org/10.1214/20-AAP1602}.

\bibitem[Welling and Teh(2011)]{welling2011sgld}
M.~Welling and Y.~W. Teh.
\newblock Bayesian learning via stochastic gradient {langevin} dynamics.
\newblock In \emph{Proceedings of the 28th International Conference on
  International Conference on Machine Learning}, ICML'11, page 681–688,
  Madison, WI, USA, 2011. Omnipress.
\newblock ISBN 9781450306195.
\newblock URL \url{https://icml.cc/2011/papers/398_icmlpaper.pdf}.

\bibitem[Younes(2000)]{younes2000convergence}
L.~Younes.
\newblock On the convergence of markovian stochastic algorithms with rapidly
  decreasing ergodicity rates.
\newblock \emph{Stochastics and Stochastics Reports}, 65, 04 2000.
\newblock \doi{10.1080/17442509908834179}.

\end{thebibliography}

\end{document}